\documentclass[a4paper,11pt]{article}
\pdfoutput=1
\usepackage{jcappub}
\usepackage{slashed}
\usepackage[section]{placeins}
\usepackage[utf8]{inputenc}

\newcommand{\Trh}{T_\text{rh}}
\newcommand{\Tmax}{T_\text{max}}
\newcommand{\arh}{a_\text{rh}}
\newcommand{\amax}{a_\text{max}}
\newcommand{\gs}{g_\star}
\newcommand{\gss}{g_{\star s}}
\newcommand{\mdm}{m_\text{DM}}
\newcommand{\Tprh}{{T'_\text{rh}}}
\newcommand{\Cr}{\mathcal{C}_\rho}
\newcommand{\Cn}{\mathcal{C}_n}

\title{Gravitational SIMPs}

\author[a,\, b]{Basabendu Barman}
\author[b]{and Nicolás Bernal}
\affiliation[a]{Department of Physics, IIT Guwahati, Guwahati 781039, India}
\affiliation[b]{Centro de Investigaciones, Universidad Antonio Nariño\\
Carrera 3 este \# 47A-15, Bogotá, Colombia}
\emailAdd{bb1988@iitg.ac.in}
\emailAdd{nicolas.bernal@uan.edu.co}

\abstract{
We study the impact of thermalization and number-changing processes in the dark sector on the yield of gravitationally produced dark matter (DM).
We take into account the DM production through the $s$-channel exchange of a massless graviton both from the scattering of inflatons during the reheating era, and from the Standard Model bath via the UV freeze-in mechanism. By considering the DM to be a scalar, a fermion, and a vector boson we show, in a model-independent way, that DM self-interaction gives rise to a larger viable parameter space by allowing lower reheating temperature to be compatible with Planck observed relic abundance. As an example, we also discuss our findings in the context of the $\mathbb{Z}_2$-symmetric scalar singlet DM model.
}

\begin{document}
\begin{flushright}
    PI/UAN-2021-688FT
\end{flushright}

\maketitle

\section{Introduction}\label{sec:intro} 

The existence of dark matter (DM) has been proven beyond any doubt from several astrophysical~\cite{Zwicky:1933gu, Zwicky:1937zza, Rubin:1970zza, Clowe:2006eq} and cosmological~\cite{Hu:2001bc, Aghanim:2018eyx} evidences (for a review, see, e.g. Refs.~\cite{Jungman:1995df, Bertone:2004pz, Feng:2010gw}). Its fundamental nature, however, still remains elusive. As it is already established from observations, DM as a fundamental particle has to be electrically neutral and stable at the scale of lifetime of the Universe. The measurement of the anisotropy in the cosmic microwave background radiation (CMB) provides the most precise measurement of the DM relic density, usually expressed as $\Omega_\text{DM}h^2 \simeq 0.12$~\cite{Aghanim:2018eyx}, which is an important constraint to abide by. Since the Standard Model (SM) of particle physics fails to offer a viable candidate, one has to look beyond the realms of the SM to explain the particle DM. 

The absence of any significant excess in direct search experiments like XENON~\cite{Aprile:2018dbl} have cornered DM as a weakly interacting massive particle (WIMP) (see, e.g., Refs.~\cite{Arcadi:2017kky, Roszkowski:2017nbc}) which, by far, is the most popular DM candidate that can be found in myriad extensions to the SM. Here one assumes that the DM particles were in thermal equilibrium in the early Universe due to non-negligible coupling with the visible sector. The DM abundance then freezes out once the interaction rate falls out of equilibrium as the Universe expands and cools down. It is also quite possible that DM particles never attain thermal equilibrium because either they are too massive or they have very feeble couplings with the SM. In such a case DM is dubbed as a feebly interacting massive particle (FIMP). 
Such nonthermal DM yield freezes in to provide the observed DM abundance.
Depending on whether the DM interaction with the visible sector is renormalizable or non-renormalizable, freeze-in in principle can be of two kinds: $i)$~infrared (IR) where the DM abundance becomes important at a low temperature~\cite{Hall:2009bx, Chu:2011be, Bernal:2017kxu, Duch:2017khv, Biswas:2018aib, Chakraborti:2019ohe, Barman:2019lvm, Heeba:2019jho}, or $ii)$ ultra-violate (UV) where the DM genesis takes place at the highest temperature achieved by the thermal bath~\cite{Hall:2009bx, Elahi:2014fsa, Chen:2017kvz, Bernal:2019mhf, Biswas:2019iqm, Barman:2020plp, Barman:2020ifq, Barman:2021tgt}. This temperature can be the reheating temperature in the case of an instantaneous and complete inflaton decay, but can also be much larger if the decay of the inflaton is non-instantaneous~\cite{Giudice:1999fb, Giudice:2000ex}. During inflationary reheating the DM can also be produced from the decay of the inflaton itself, either directly~\cite{Ellis:2015jpg, Garcia:2017tuj, Dudas:2017rpa, Garcia:2020eof} or radiatively~\cite{Kaneta:2019zgw}. On top of decay, pair annihilation of inflatons mediated by gravity can also lead to sufficient DM abundance during inflation~\cite{Ema:2018ucl, Mambrini:2021zpp, Bernal:2021kaj}.

The irreducible gravitational interaction can lead to DM production which, in general, can be dubbed as the gravitational production.
Several mechanisms for the gravitational particle production exist. For example, particles are generated due to time variations in the background metric~\cite{Parker:1969au, Birrell:1982ix, Kuzmin:1998kk, Kuzmin:1999zk, Chung:2001cb}, with a maximal efficiency when the particle mass $m\simeq H_I$~\cite{Garny:2015sjg}, where $H_I$ denotes the Hubble scale at the end of inflation.
Production of superheavy (quasi-) stable particles with masses $m\gg H_I$ during the transition from the inflationary phase to either a matter- or a radiation-dominated phase as a result of the expansion of the background spacetime has been studied in the context of a scalar field conformally coupled to gravity~\cite{Chung:1998zb, Chung:2004nh}, and for a wide range of models with minimal and/or conformal coupling to gravity, where it has been shown such particles can constitute a considerable fraction of the DM~\cite{Kuzmin:1998kk}. The production of purely gravitational DM due to expansion of the Universe has been extensively discussed in Refs.~\cite{Tang:2016vch, Ema:2018ucl, Hashiba:2018tbu}, while DM production through $s$-channel graviton exchange has been studied in Refs.~\cite{Garny:2015sjg, Tang:2016vch, Garny:2017kha, Tang:2017hvq}. The latter may as well be regarded as gravitational particle creation since no direct or indirect interaction, apart from gravity, between the DM and the SM field is considered. It has been pointed out that (small) oscillations of the Hubble parameter or the scale factor caused by inflaton oscillation induces an effective coupling between the inflaton and a (massive) particle of arbitrary spin, leading to particle creation in the inflaton oscillation regime~\cite{Ema:2015dka, Ema:2016hlw}. Such a particle production mechanism can also be interpreted as the decay or annihilation of the inflaton through the gravitational interaction. In the present context, however, the DM particle production takes place via gravitational interaction which is described in terms of coupling of the energy-momentum tensor $T^{\mu\nu}$ to the metric perturbation $h^{\mu\nu}$~\cite{Donoghue:1994dn, Choi:1994ax, Holstein:2006bh} where the latter is identified as the quantum field for the spin-2 massless graviton. This gives rise to interactions of all matter fields with the graviton and subsequent pair production of DM particles. Note that, the present framework differs from those where non-minimal couplings are introduced~\cite{Ren:2014mta, Cata:2016epa}, and also from Refs.~\cite{Chung:1998ua, Kuzmin:1998kk} which rely on the dynamics of quantum field theory on curved background spacetimes.


The DM particles, on the other hand, can have strong self-interactions among themselves leading to the strongly interacting massive particle (SIMP) paradigm that can have significant impact on the resulting phenomenology~\cite{Hochberg:2014dra}. The long-standing puzzles of the collisionless cold DM paradigm can be addressed by invoking these interactions. Two such examples are the `core vs cusp'~\cite{Flores:1994gz, Oh:2010mc, Walker:2011zu} and the `too big to fail'~\cite{BoylanKolchin:2011de} problems which can be alleviated if at the scale of galaxies there exists a large self-scattering cross section. DM self-interaction can also accommodate the observed diversity in galaxy rotation curves with the same circular velocity~\cite{Kaplinghat:2015aga, Kamada:2016euw, Tulin:2017ara}.
Although a large self-scattering cross section is constrained by Bullet cluster spherical halo shapes~\cite{Markevitch:2003at, Clowe:2003tk, Randall:2007ph}, it can lead to distinct signatures in galaxies and galaxy clusters, such as the offset of the dark matter subhalo from the galaxy centre as hinted in Abell-3827~\cite{Kahlhoefer:2015vua}. However, the dynamics within the dark sector can be non-trivial in case there is a $N$-to-$N'$ number changing process involving DM particles with $N > N' \geq 2$. Typically such a number-changing process corresponds to 3-to-2 scenario~\cite{Carlson:1992fn, Hochberg:2014dra, Bernal:2015bla, Bernal:2015ova, Bhattacharya:2019mmy} but an unavoidable 4-to-2 process may dominate depending on the stabilizing symmetry of the dark sector~\cite{Bernal:2015xba, Bernal:2017mqb, Bernal:2018hjm, Bernal:2018ins}. If the number-changing processes within the dark sector reach equilibrium, DM forms a thermal bath with a temperature $T'$ in general different from $T$ of the SM. Such number-changing interactions naturally affect the DM abundance once the DM is produced.

Motivated from these, in this work we investigate the effect of thermalization due to number-changing processes within the dark sector on the abundance of gravitationally produced DM, where gravity minimally couples to all matter particles through the energy-momentum tensor.%
\footnote{The effect of strong self-interactions has also been studied in the case of conformal~\cite{Redi:2020ffc, Gross:2020zam, Garani:2021zrr} and non-minimally coupled~\cite{Fairbairn:2018bsw} dark sectors.}
We focus on the DM production $a)$ during the reheating epoch when the inflatons dominate the energy density of the Universe, and $b)$ through the UV freeze-in mechanism via 2-to-2 annihilation of the SM states, both mediated by the $s$-channel exchange of massless gravitons.%
\footnote{Two comments are in order: First, one could also conceive scenarios where gravity and another portal are effective, see e.g., Refs.~\cite{Chianese:2020yjo, Chianese:2020khl, Gondolo:2020uqv, Bernal:2020bjf, Bernal:2021akf}.
Second, the gravitational production can be enhanced in scenarios with extra dimensions, see e.g., Refs.~\cite{Lee:2013bua, Lee:2014caa, Han:2015cty, Rueter:2017nbk, Kim:2017mtc, Rizzo:2018ntg, Carrillo-Monteverde:2018phy, Kim:2018xsp, Rizzo:2018joy, Goudelis:2018xqi, Brax:2019koq, Folgado:2019sgz, Goyal:2019vsw, Folgado:2019gie, Kang:2020huh, Chivukula:2020hvi, Kang:2020yul, Kang:2020afi, Bernal:2020fvw, Bernal:2020yqg}.}
In a model-independent way, considering the DM to be a scalar, a fermion or a vector boson, we show that thermalization broadens the viable parameter space allowing lower reheating temperature compatible with the DM relic abundance. We also show that, for the fermionic case, DM production from the thermal bath (mediated by gravity) could dominate for $\Tmax/\Trh\lesssim 20$ once self-interaction is invoked. In the later part of the analysis, as an example, we demonstrate in detail the consequences of thermalization on the gravitationally produced scalar singlet DM (SSDM)~\cite{Silveira:1985rk, McDonald:1993ex, Burgess:2000yq} , where we see the DM abundance typically increases on increasing self-interaction strength before non-relativistic freeze-out occurs. 

The paper is organized as follows. In section~\ref{sec:GravProd} we review in detail the DM gravitational production, i.e., via 2-to-2 annihilations of inflatons and the SM particles mediated by $s$-channel exchange of gravitons. In section~\ref{sec:sidm} the role of DM self-interaction is presented in a model independent way emphasizing its impact on the DM relic abundance.
In section~\ref{sec:ssdm} we briefly review the singlet scalar DM model as an example of gravitational DM with sizeable self-interactions. Finally, we conclude in section~\ref{sec:con}.

\section{Dark Matter Gravitational Production} \label{sec:GravProd}
Gravity is the only interaction that is guaranteed to mediate between the DM and the visible sector. In that context, as mentioned in the introduction, DM can be produced by scattering of inflatons or SM particles, mediated by $s$-channel exchange of gravitons. In such a scenario, the Boltzmann equation (BEQ) governing the evolution of the DM number density $n$ is given by 
\begin{equation} \label{eq:BE0}
    \frac{dn}{dt} + 3H\,n = \gamma\,,
\end{equation}
where $\gamma$ corresponds to the DM production rate density and $H$ to the Hubble expansion rate given by $H^2 = (\rho_R + \rho_\phi)/(3M_P^2)$, with $M_P$ as the reduced Planck mass. The evolution of the inflaton and SM radiation energy densities ($\rho_\phi$ and $\rho_R$, respectively) can be tracked via the set of BEQs
\begin{align}
    &\frac{d\rho_\phi}{dt} + 3H\, \rho_\phi = - \Gamma_\phi\, \rho_\phi\,, \label{eq:BErhop}\\
    &\frac{d\rho_R}{dt} + 4H\, \rho_R = + \Gamma_\phi\, \rho_\phi\,, \label{eq:BErhoR}
\end{align}
where $\Gamma_\phi$ is the total inflaton decay width.
We note that, by construction, the inflaton does not decay into DM particles as DM is assumed to only have gravitational interactions with the visible sector.%
\footnote{Here we are assuming that in the reheating era the inflaton energy density scales like non-relativistic matter.
This can be achieved in quadratic inflation, or in non-minimally coupled scenarios like in the case of Starobinsky inflation.}

\begin{figure}
    \def\sepf{0.50}
	\centering
	\includegraphics[scale=\sepf]{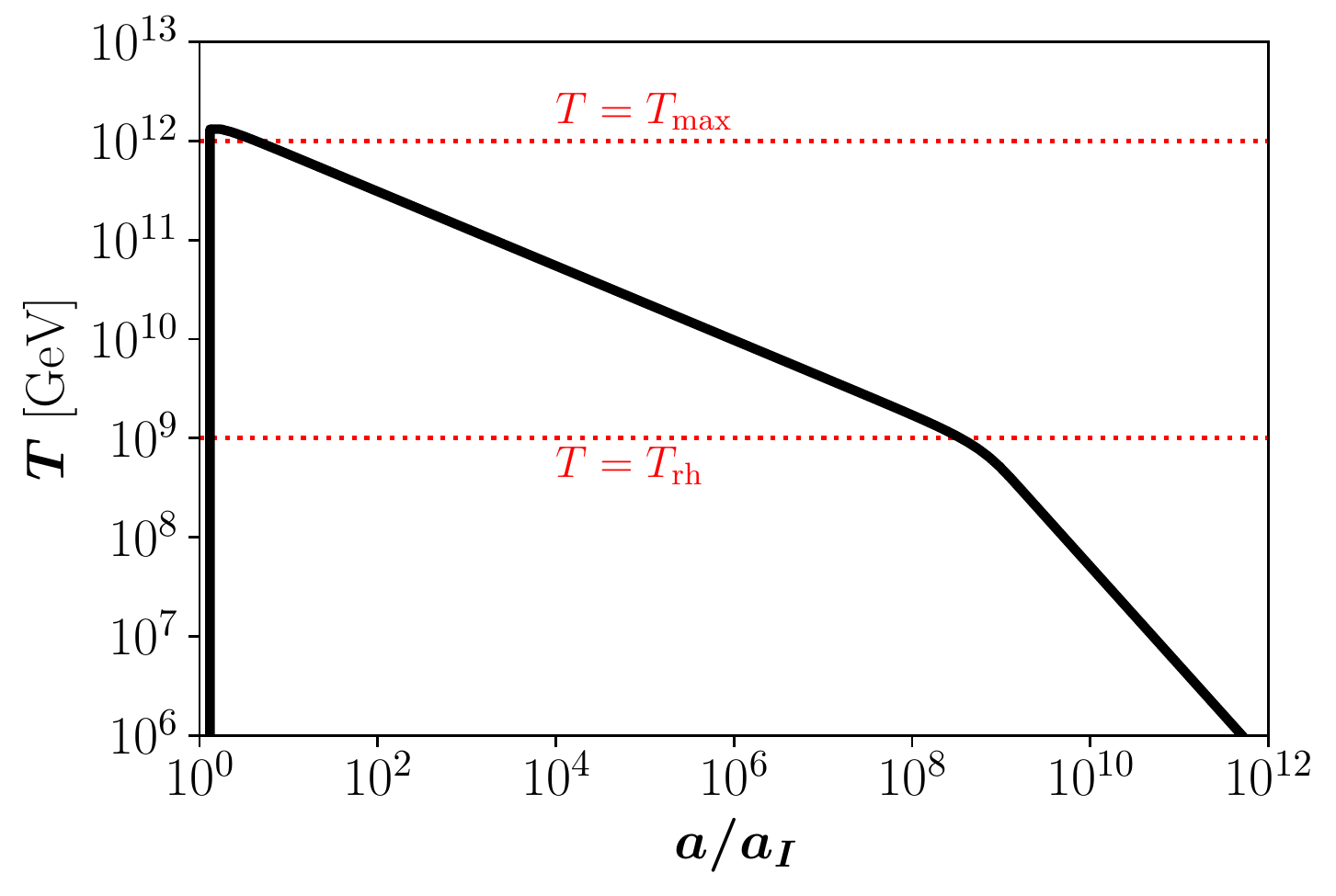}
    \caption{Evolution of the SM temperature as a function of the scale factor for $\Tmax = 10^{12}$~GeV and $\Trh = 10^9$~GeV.}
	\label{fig:Tevolution}
\end{figure}

The photon temperature can be extracted from the SM energy density $\rho_R = \frac{\pi^2}{30}\, \gs\, T^4$, where $\gs(T)$ corresponds to the SM relativistic degrees of freedom contributing to the energy density~\cite{Drees:2015exa}. We note that the SM particles do not necessarily thermalize instantaneously, and thus the decay products could be initially distributed with smaller occupation numbers and harder momenta~\cite{Harigaya:2013vwa, Ellis:2015jpg, Garcia:2018wtq}. However, here we assume that the SM thermalizes rapidly. The reheating temperature $\Trh$ can then be defined as the temperature at which the equality $H(\Trh) = \Gamma_\phi$ holds, and is given by
\begin{equation}
    \Trh^2 = \frac{3}{\pi} \sqrt{\frac{10}{\gs}}\, M_P\, \Gamma_\phi\,.\label{eq:trh}
\end{equation}
It is also a proxy of the onset of the radiation domination era.
As the inflaton does not decay instantaneously, the SM bath temperature may rise to a value $\Tmax$ much higher than $\Trh$. The value of $\Tmax$ depends both on $\Gamma_\phi$ and on the inflationary scale, and is given in Eq.~\eqref{eq:Tmax}. During the heating era, i.e., while $\Tmax > T > \Trh$, the inflaton decays into SM radiation. As a result, the SM energy density does not scale as free radiation but rather like $\rho_R(a) \propto a^{-3/2}$, with $a$ as the scale factor. 
Figure~\ref{fig:Tevolution} shows the SM temperature evolution as a function of the scale factor obtained from the numerical solution of Eqs.~\eqref{eq:BErhop} and~\eqref{eq:BErhoR},\footnote{It is useful to make the variable transformation $\Phi \equiv \rho_\phi a^3$ and $R \equiv \rho_R a^4$ to solve the system of BEQs~\cite{Giudice:2000ex}.} for $\Tmax = 10^{12}$~GeV and $\Trh = 10^9$~GeV shown by the red dashed lines. During reheating $T \propto a^{-3/8}$, whereas once the reheating is completed, $T \propto a^{-1}$ as the inflaton has already decayed and hence the entropy injection has ceased.%
\footnote{We note that the scaling $T \propto a^{-3/8}$ characterizes scenarios where a non-relativistic particle that dominates the total energy density (here the inflaton) decays into SM states.}

\subsection{Inflaton Scatterings}\label{sec:inf-scat}
The total energy density of the Universe is dominated by inflatons during reheating. Taking into account that $T \propto a^{-3/8}$ and the inflaton energy density scales as non-relativistic matter,
\begin{equation}
    \rho_\phi(T) = \frac{\pi^2\, \gs}{30}\, \frac{T^8}{\Trh^4}\,,
\end{equation}
where $\rho_\phi(\Trh) = \rho_R(\Trh)$ is assumed. Therefore, the Hubble parameter in an inflaton-dominated Universe takes the form 
\begin{equation}
    H(T) = \frac{\pi}{3} \sqrt{\frac{\gs}{10}}\, \frac{T^4}{M_P\, \Trh^2}\,.
\end{equation}

During reheating, the whole observed DM abundance can be generated via 2-to-2 annihilations of inflatons, mediated by the $s$-channel exchange of gravitons. The interaction rate density for DM production out of non-relativistic inflatons then reads~\cite{Mambrini:2021zpp, Bernal:2021kaj}
\begin{equation} \label{eq:gamma}
    \gamma = \frac{\rho_\phi^2}{4096\pi\, M_P^4}\, f\left(\frac{\mdm}{m_\phi}\right)
    = \frac{\pi^3\, \gs^2}{3686400}\, \frac{T^{16}}{M_P^4\, \Trh^8}\, f\left(\frac{\mdm}{m_\phi}\right),
\end{equation}
where $\mdm$ and $m_\phi$ are the masses of the DM and the inflaton, respectively, and we have defined
\begin{equation}
f(x) \equiv
    \begin{cases}
        \left(x^{2}+2\right)^{2}\sqrt{1-x^{2}} & \mbox{for real scalars},\\[8pt]
        x^{2}\left(1-x^{2}\right)^{3/2} & \mbox{for Dirac fermions},\\[8pt]
        \frac{1}{8}\sqrt{1-x^2}\left(4+4x^2+19x^4\right)&
        \mbox{for vector bosons.}
    \end{cases}
\end{equation}
For completeness, in Appendix~\ref{sec:vertex} we have gathered the relevant vertices between the graviton and particles with different spins (Tab.~\ref{tab:vert-fac}). Additionally, the details of the computation of the rates are presented in Appendix~\ref{sec:rates}.

\begin{figure}
    \def\sepf{0.50}
	\centering
	\includegraphics[scale=\sepf]{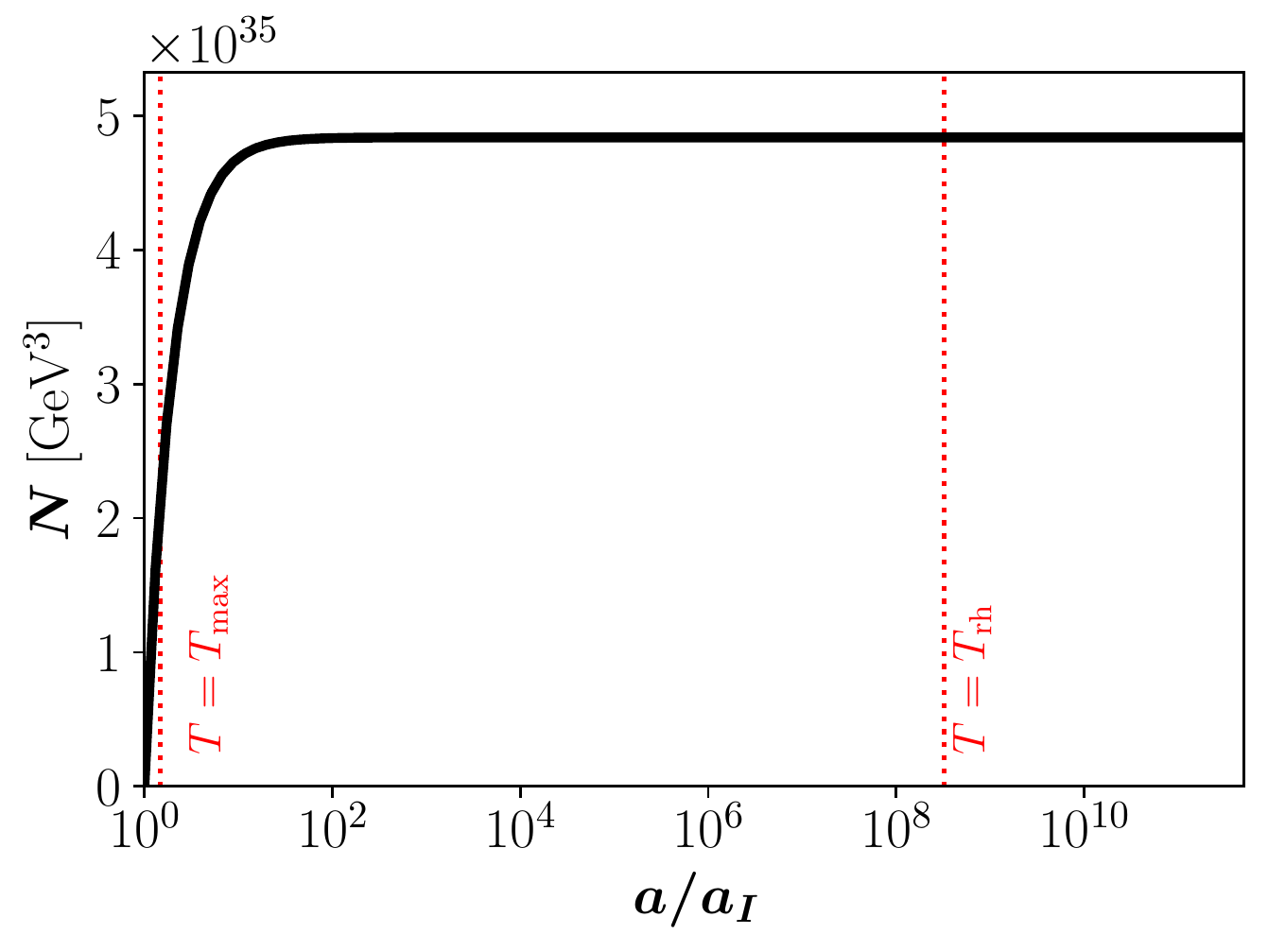}
	\includegraphics[scale=\sepf]{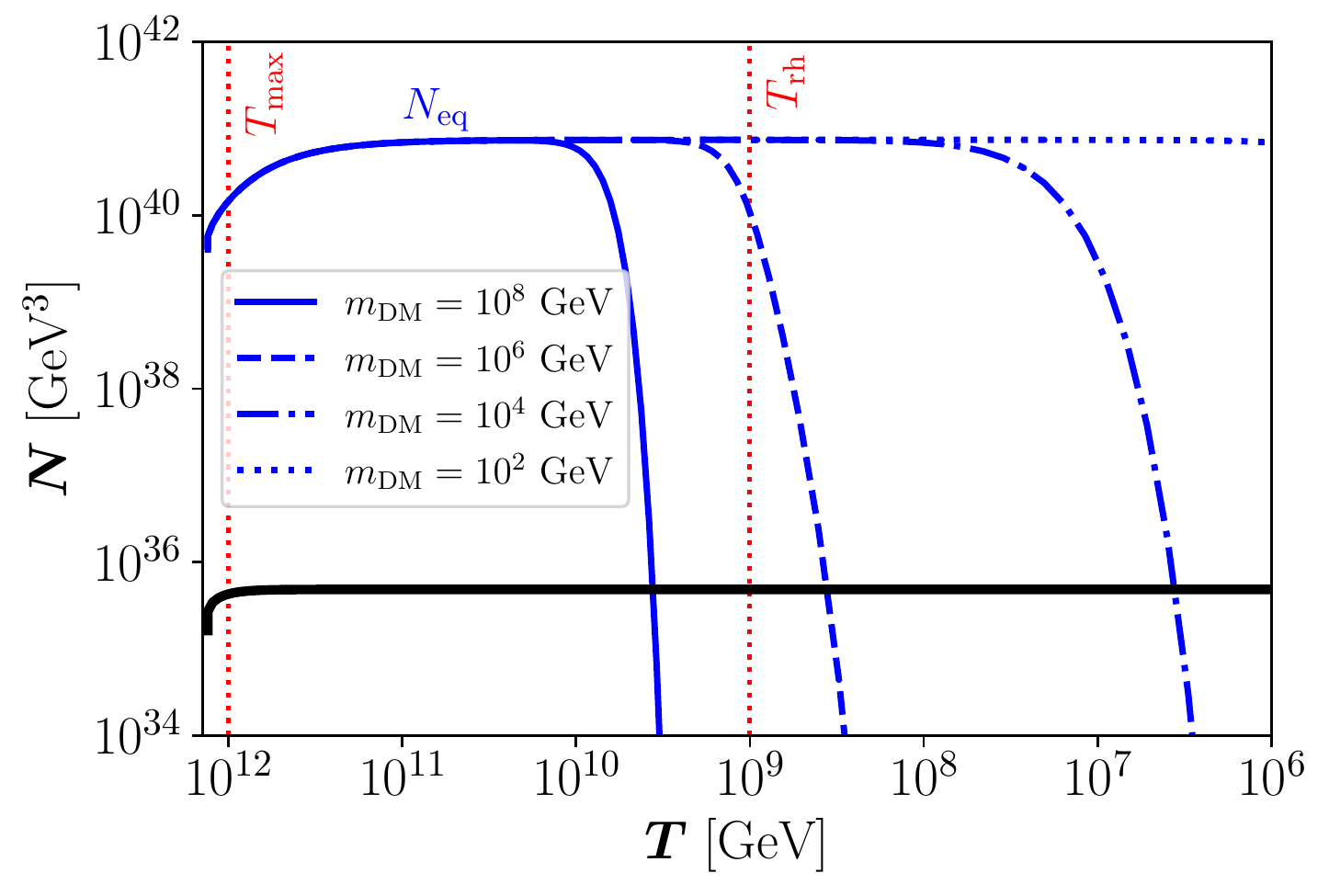}
    \caption{Evolution of the comoving DM yield $N$ as a function of the SM temperature $T$ (black solid line), for $\Tmax= 10^{12}$~GeV and $\Trh = 10^9$~GeV.
    The blue lines correspond to the equilibrium yields, for different DM masses.
    The vertical red dotted lines correspond to $T = \Tmax$ and $T = \Trh$.
    }
	\label{fig:DMevolution}
\end{figure}

The evolution of the DM number density given in Eq.~\eqref{eq:BE0} can be recasted in terms of the co-moving number density $N \equiv n\, a^3$, considering the fact that during the reheating era the SM entropy is not conserved due to the inflaton decay
\begin{equation} \label{eq:BE1}
    \frac{dN}{dT} = -\frac{8}{\pi} \sqrt{\frac{10}{\gs}}\, \frac{M_P\, \Trh^{10}}{T^{13}}\, a^3(\Trh)\, \gamma\,.
\end{equation}
This equation integrated in the range $\Tmax \geq T \geq \Trh$ gives the final DM number density. The DM yield $Y(T) \equiv n(T)/s(T)$ is defined as a function of the SM entropy density $s \equiv \frac{2\pi^2}{45}\gss\, T^3$ with $\gss(T)$ being the number of relativistic degrees of freedom contributing to the SM entropy~\cite{Drees:2015exa}. The yield $Y_0$ at the end of reheating (i.e., at $T = \Trh$) can be analytically computed and reads
\begin{equation} \label{eq:Y0}
    Y_0 = \frac{\gs^2}{81920\, \gss} \sqrt{\frac{10}{\gs}}\, \left(\frac{\Trh}{M_P}\right)^3 \left[\left(\frac{\Tmax}{\Trh}\right)^4 - 1\right] f\left(\frac{\mdm}{m_\phi}\right).
\end{equation}
Figure~\ref{fig:DMevolution} shows, with black solid lines, the evolution of the comoving DM yield $N$ as a function of the scale factor $a$ (left panel) and the SM temperature $T$ (right panel) for DM produced via inflaton scatterings during reheating. Here we assume $\Tmax= 10^{12}$~GeV, $\Trh = 10^9$~GeV shown by the red dashed vertical lines, and a scalar DM particle with mass $\mdm \ll m_\phi$. The bulk of the DM is produced promptly at the beginning of reheating, when $T \simeq \Tmax$.

\begin{figure}
    \def\sepf{0.39}
	\centering
	\includegraphics[scale=\sepf]{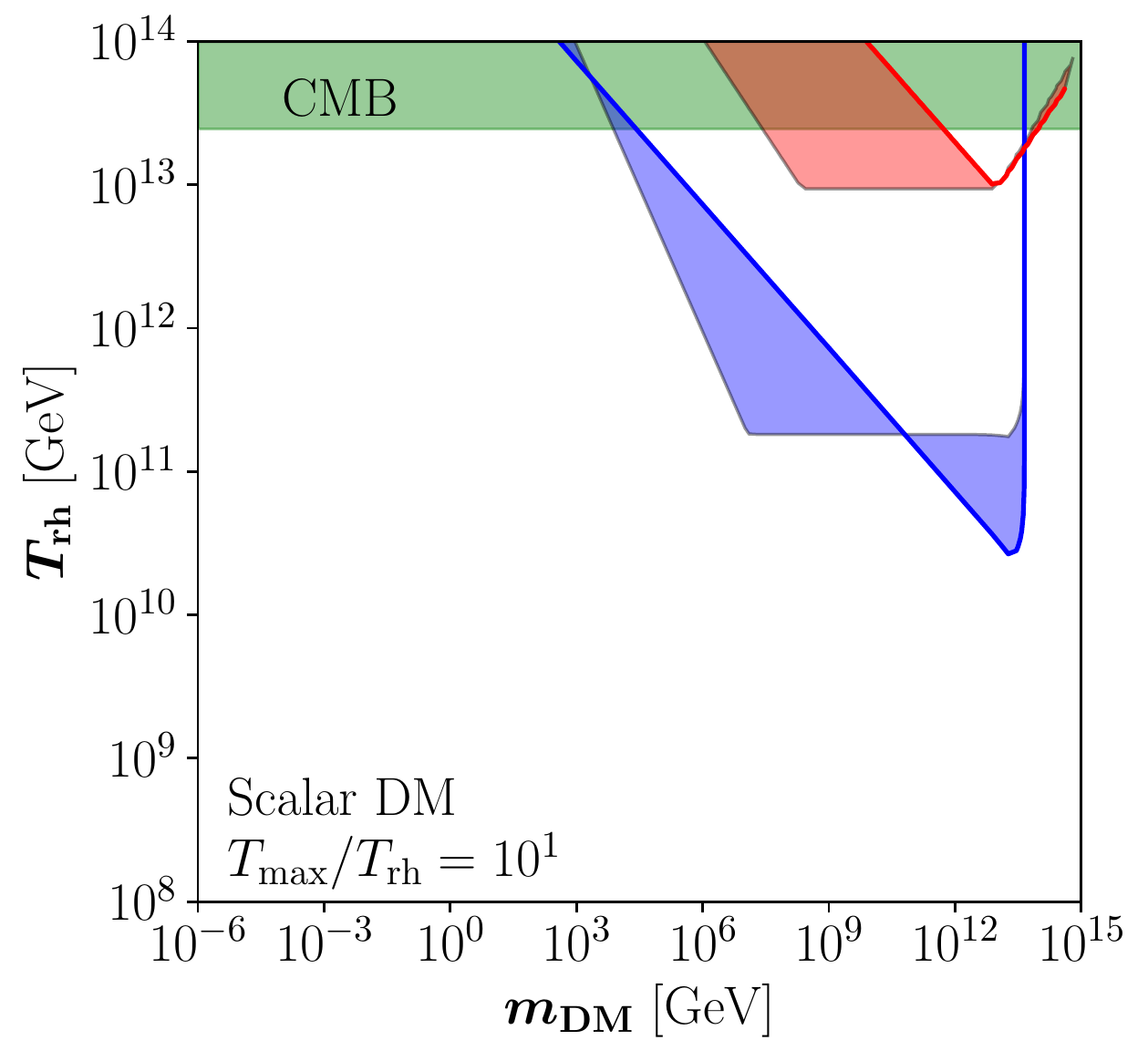}
	\includegraphics[scale=\sepf]{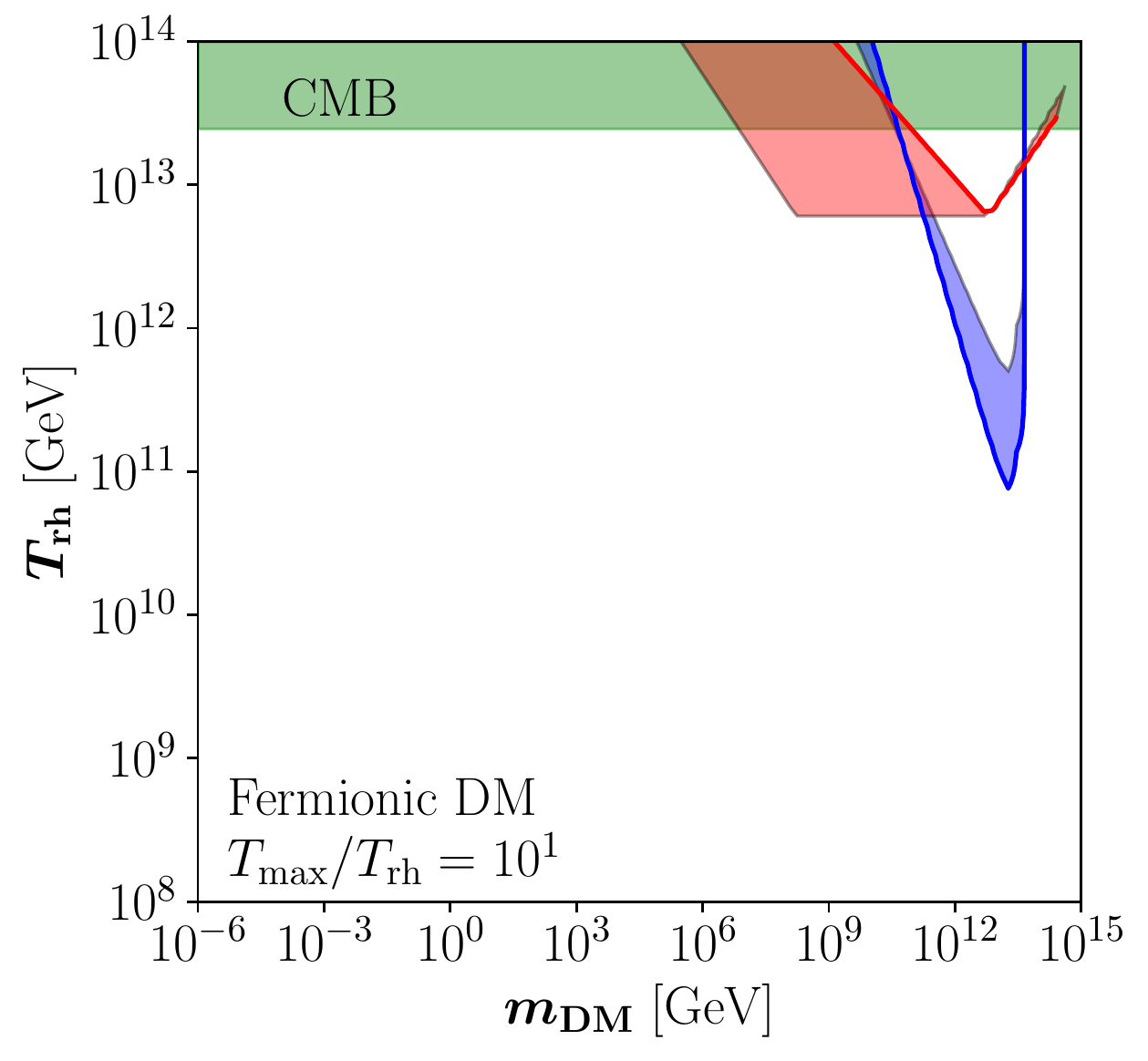}
	\includegraphics[scale=\sepf]{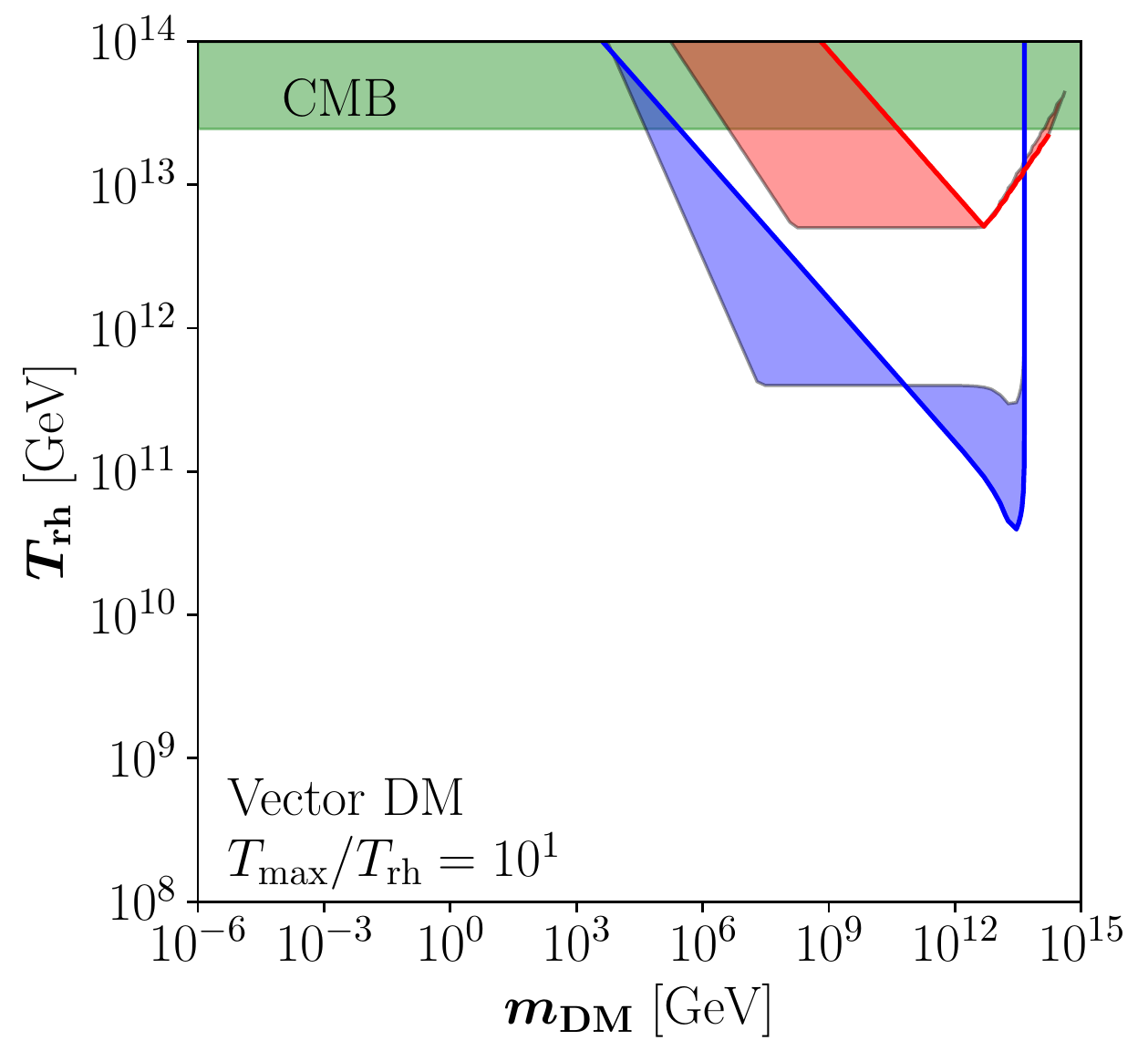}
	\includegraphics[scale=\sepf]{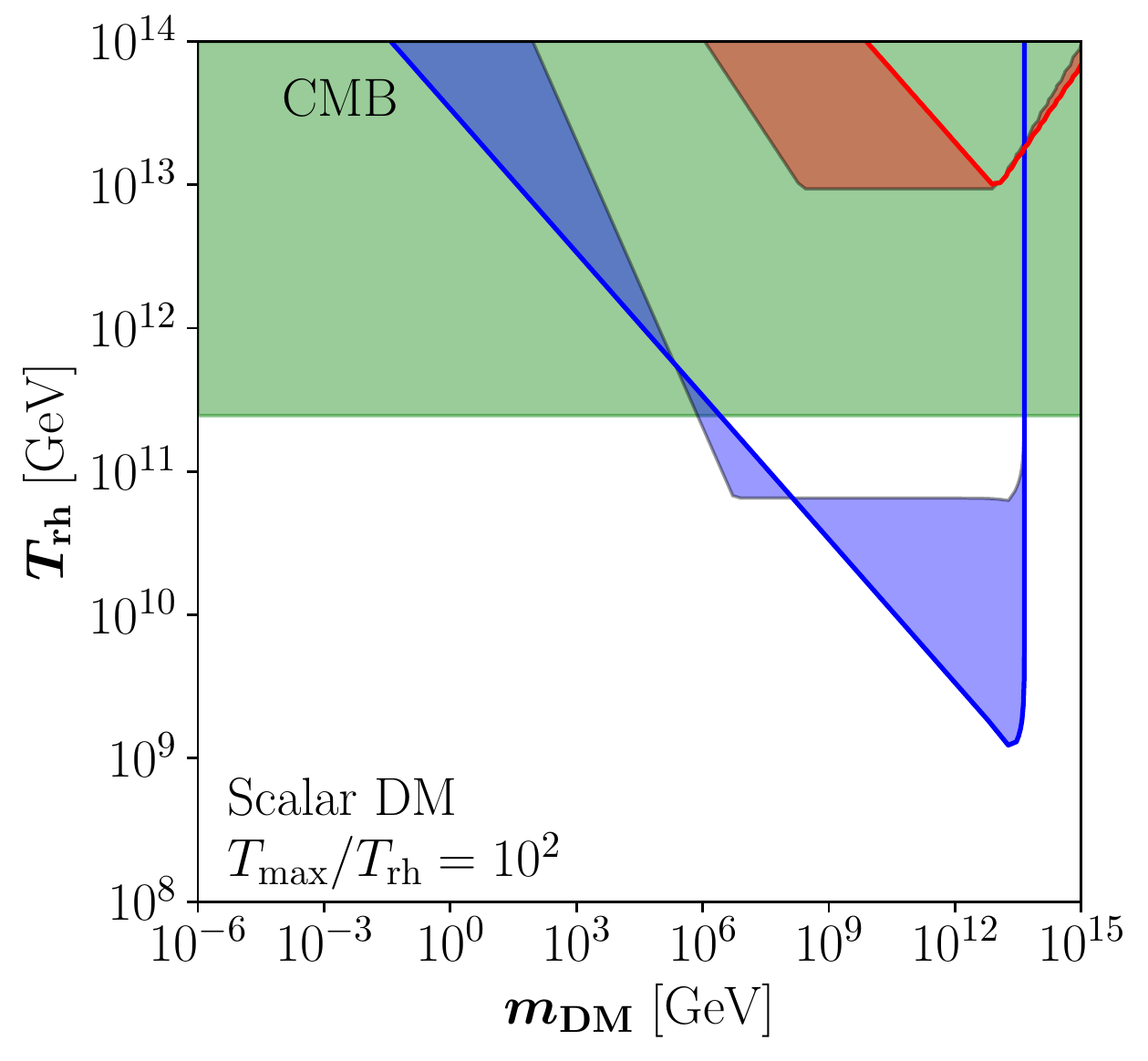}
	\includegraphics[scale=\sepf]{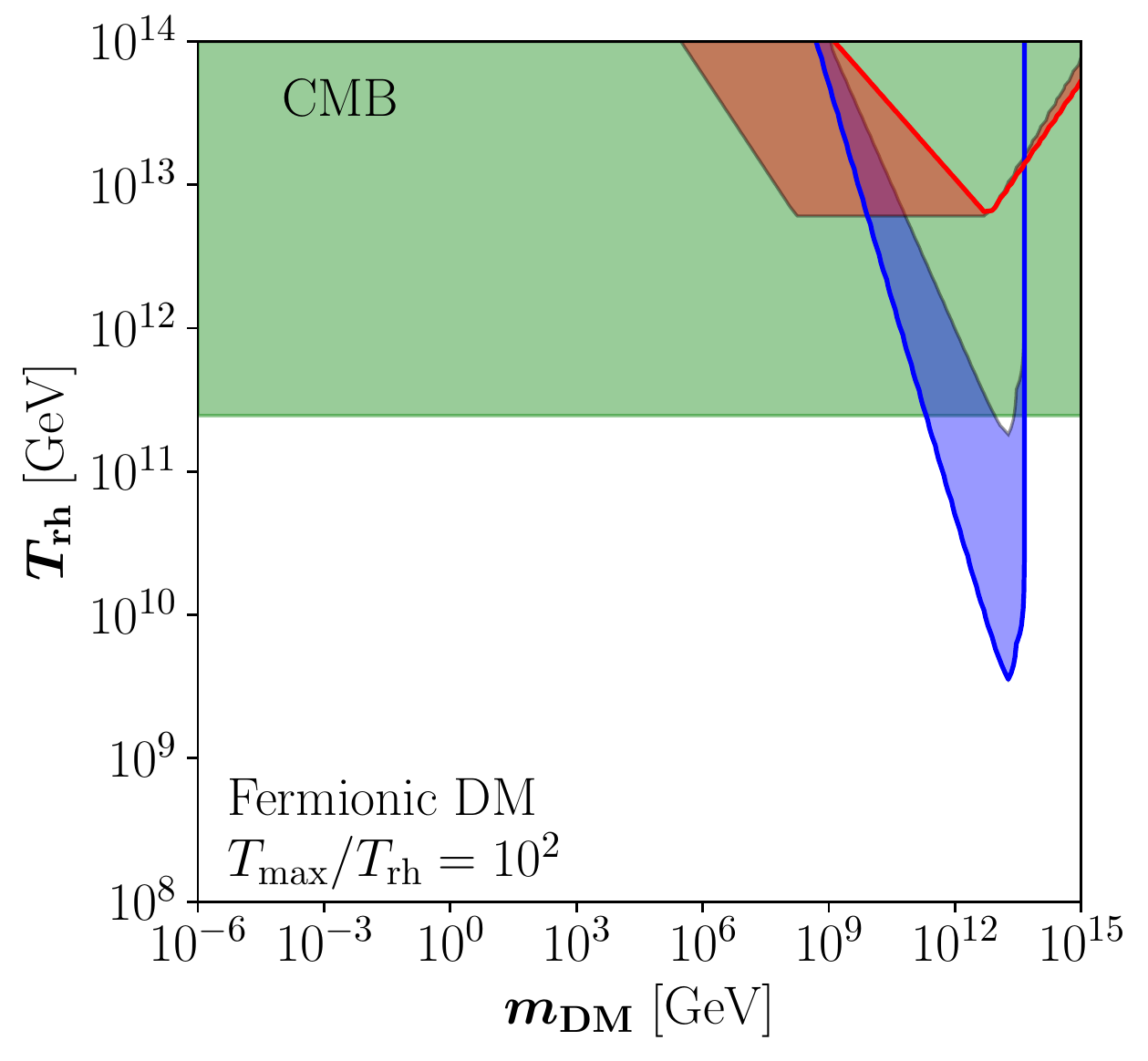}
	\includegraphics[scale=\sepf]{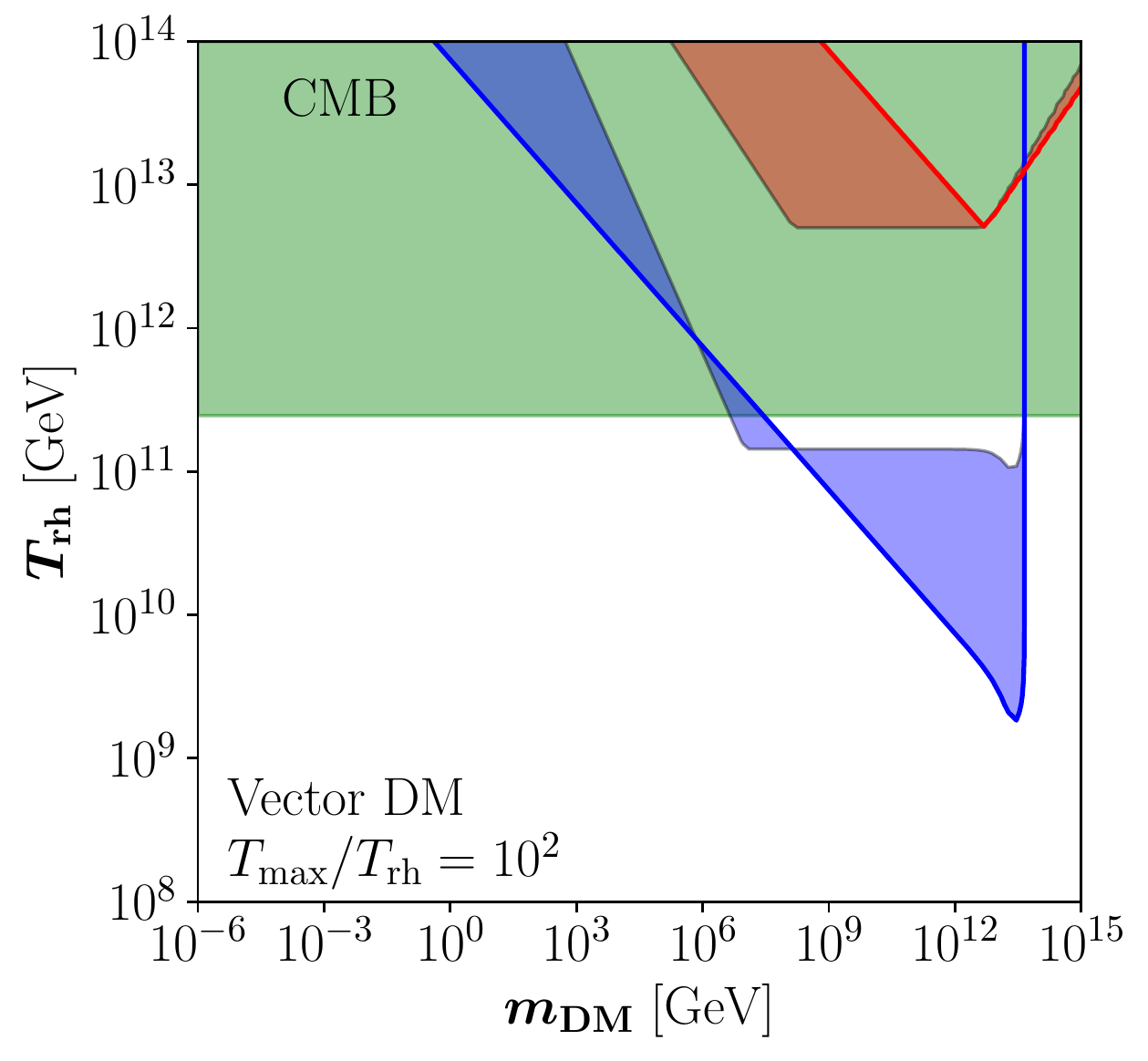}
    \caption{Parameter space that generates the observed DM abundance for scalar (left panels), fermionic (central panels) and vector (right panels) DM, for $m_\phi = 3 \times 10^{13}$~GeV, and $\Tmax/\Trh = 10^1$ (upper panels) or $\Tmax/\Trh = 10^2$ (lower panels).
    DM is produced by inflaton (blue areas) or SM (red areas) scatterings mediated by the $s$-channel exchange of gravitons.
    Blue and red lines correspond to the case without thermalization, whereas the width of the bands to the effect of thermalization and number-changing processes. Above (below) the bands DM is overabundant (underabundant)
    }
	\label{fig:DM}
\end{figure}

To match the observed DM abundance $\Omega_\text{DM} h^2 \simeq 0.12$, the DM yield has to be fixed so that $\mdm Y_0 = \Omega_\text{DM} h^2 \frac{1}{s_0} \frac{\rho_c}{h^2} \simeq 4.3 \times 10^{-10}$~GeV, where $\rho_c \simeq 1.1 \times 10^{-5} h^2$~GeV/cm$^3$ is the critical energy density and $s_0 \simeq 2.9 \times 10^3$~cm$^{-3}$ is the entropy density at present~\cite{Aghanim:2018eyx}. Figure~\ref{fig:DM} shows the regions of the parameter space that generate the observed DM abundance for scalar (left panels), fermionic (central panels) and vector (right panels) DM. We consider $m_\phi = 3 \times 10^{13}$~GeV, and $\Tmax/\Trh = 10^1$ (upper panels) or $\Tmax/\Trh = 10^2$ (lower panels). Blue {\it lines} correspond to the DM production via inflaton annihilation as in Eq.~\eqref{eq:Y0}. The strong cuts at $\mdm = m_\phi$ correspond to the kinematical production threshold. Furthermore, regions above the lines generate a DM overdensity, overclosing the Universe. Large values of the ratio $r \equiv \Tmax/\Trh$ increase the DM production as $Y_0\sim \Trh^3\, r^4$ following Eq.~\eqref{eq:Y0} for a fixed DM mass and reheating temperature. As a result, for a given $r$, a lower $\Trh$ is required to obtain the correct DM abundance. Large values of $\Tmax/\Trh$ are also in conflict with the scale of inflation $H_I^\text{CMB} \leq 2.5 \times 10^{-5}~M_P$~\cite{Akrami:2018odb} coming from the CMB observations by the Planck satellite. The upper green bands (labeled CMB) correspond to the reheating temperatures that are in tension with the CMB observations, as  derived in Eq.~\eqref{eq:boundHI} of Appendix~\ref{sec:Trh}. It is interesting to note that, if $\Tmax/\Trh \gtrsim 3 \times 10^5$, the whole parameter space where DM can be produced via the scattering of inflatons is in tension with the upper bound of the inflationary scale. Finally, we note that DM lighter than few keV is in tension with the Lyman-$\alpha$ bound~\cite{Irsic:2017ixq, Ballesteros:2020adh, DEramo:2020gpr} if produced via SM scatterings. Alternatively, if it is produced out of inflaton scatterings, this constraint can be eased as discussed in Appendix~\ref{app:lyman}.

\subsection{SM Scatterings} \label{sec:uv-fi}
Apart from the inflaton scattering, the DM is also gravitationally produced from the UV freeze-in mechanism via 2-to-2 annihilation of the SM particles mediated by $s$-channel exchange of gravitons. The interaction rate density for such a process reads~\cite{Garny:2015sjg, Tang:2017hvq, Garny:2017kha, Bernal:2018qlk}
\begin{equation}
    \gamma(T) = \alpha\, \frac{T^8}{M_P^4}\,,
\end{equation}
with $\alpha \simeq 1.9\times 10^{-4}$ (real scalar DM), $\alpha \simeq 1.1\times 10^{-3}$ (Dirac DM) or  $\alpha \simeq 2.3\times 10^{-3}$ (vector DM). In the sudden decay approximation for the inflaton, Eq.~\eqref{eq:BE0} can be recasted as
\begin{equation}
    \frac{dY}{dT} = - \frac{\gamma(T)}{H(T)\, T\, s(T)}\,,
\end{equation}
where $H(T) = \frac{\pi}{3} \sqrt{\frac{\gs}{10}} \frac{T^2}{M_P}$ in a radiation-dominated Universe. For $T \ll \Trh$, one can analytically obtain 
\begin{equation} \label{eq:FIlight}
    Y_0 = \frac{45\, \alpha}{2\pi^3\, \gss} \sqrt{\frac{10}{\gs}} \left(\frac{\Trh}{M_P}\right)^3,
\end{equation}
in the case $\mdm \ll \Trh$. Instead, if the produced particle is heavier than the reheating temperature (but still below $\Tmax$), it cannot be generated after but during reheating. In that case, the yield can be computed by integrating Eq.~\eqref{eq:BE1} in the range $\Tmax \geq T \geq \mdm$:
\begin{equation} \label{eq:FIheavy}
    Y_0 = \frac{45\, \alpha}{2\pi^3\, \gss} \sqrt{\frac{10}{\gs}} \frac{\Trh^7}{M_P^3\, \mdm^4}\,.
\end{equation}
Away from the instantaneous decay approximation, the DM yield is only boosted by a small factor of order $\mathcal{O}(1)$ for an inflaton behaving as non-relativistic matter~\cite{Garcia:2017tuj, Bernal:2019mhf}.%
\footnote{We note that big boost factors can appear when considering nonthermal effects~\cite{Garcia:2018wtq}, or expansion eras dominated by a fluid component stiffer than radiation~\cite{Bernal:2019mhf, Bernal:2020bfj}.}

Figure~\ref{fig:DM} also shows, this time with red {\it lines}, the parameter space reproducing the observed DM abundance via the UV freeze-in, i.e., by annihilation of SM particles mediated by gravitons. Two slopes are clearly visible: DM lighter than $\mdm \lesssim 10^{13}$~GeV is mainly produced at $T \simeq \Trh$ and its yield is given by Eq.~\eqref{eq:FIlight}, whereas in the opposite case ($\mdm \gtrsim 10^{13}$~GeV) the bulk of DM is generated at $T \simeq \mdm$ and the corresponding yield is given by Eq.~\eqref{eq:FIheavy}. Again, regions above the lines generate a DM overdensity, overclosing the Universe. We emphasize, even if this channel is largely independent on $\Tmax$, the bound on the inflationary scale makes it viable only if $\Tmax/\Trh \lesssim 30$. Finally, we note that this production mechanism is typically subdominant with respect to that due to the scattering of inflatons. However, it can dominate if $\mdm > m_\phi$ and $\Tmax/\Trh \lesssim 20$, where the production by inflaton annihilation is kinematically forbidden. Before moving on, we would like to clarify that the gravitational production of DM at the end of inflation due to time varying background metric has negligible contribution in the present framework compared to the production due to inflaton annihilation via $s$-channel graviton mediation. As it has been pointed out in Refs.~\cite{Garny:2015sjg, Garny:2017kha}, the gravitational production due to metric fluctuations is most relevant for $m_\text{DM}\simeq H_I$, and depending on the efficiency of reheating $\gamma_\text{rh}=\sqrt{\Gamma/H_I}$ can become important only within a relatively narrow window: $m_\text{DM}\sim 10^{-10}-10^{-7}~M_P$. Although it could dominant over the UV freeze-in (which has also been shown in Ref.~\cite{Ema:2018ucl}), it is always sub-dominant with respect to the production via inflaton scattering.

\section{Self-interacting Dark Matter}\label{sec:sidm} 
Even in the present scenario, where gravity is the only portal mediating between the visible and the dark sectors, DM can feature sizeable (i.e., non-gravitational) self-interactions. In that case, thermalization within the dark sector can occur inducing a potentially strong impact on the DM abundance~\cite{Chu:2013jja, Bernal:2015ova, Bernal:2015xba, Bernal:2017mqb, Falkowski:2017uya, Herms:2018ajr, Heeba:2018wtf, Mondino:2020lsc, Bernal:2020gzm, March-Russell:2020nun, Bernal:2020kse}.

\begin{figure}[t!]
    \def\sepf{0.50}
	\centering
	\includegraphics[scale=\sepf]{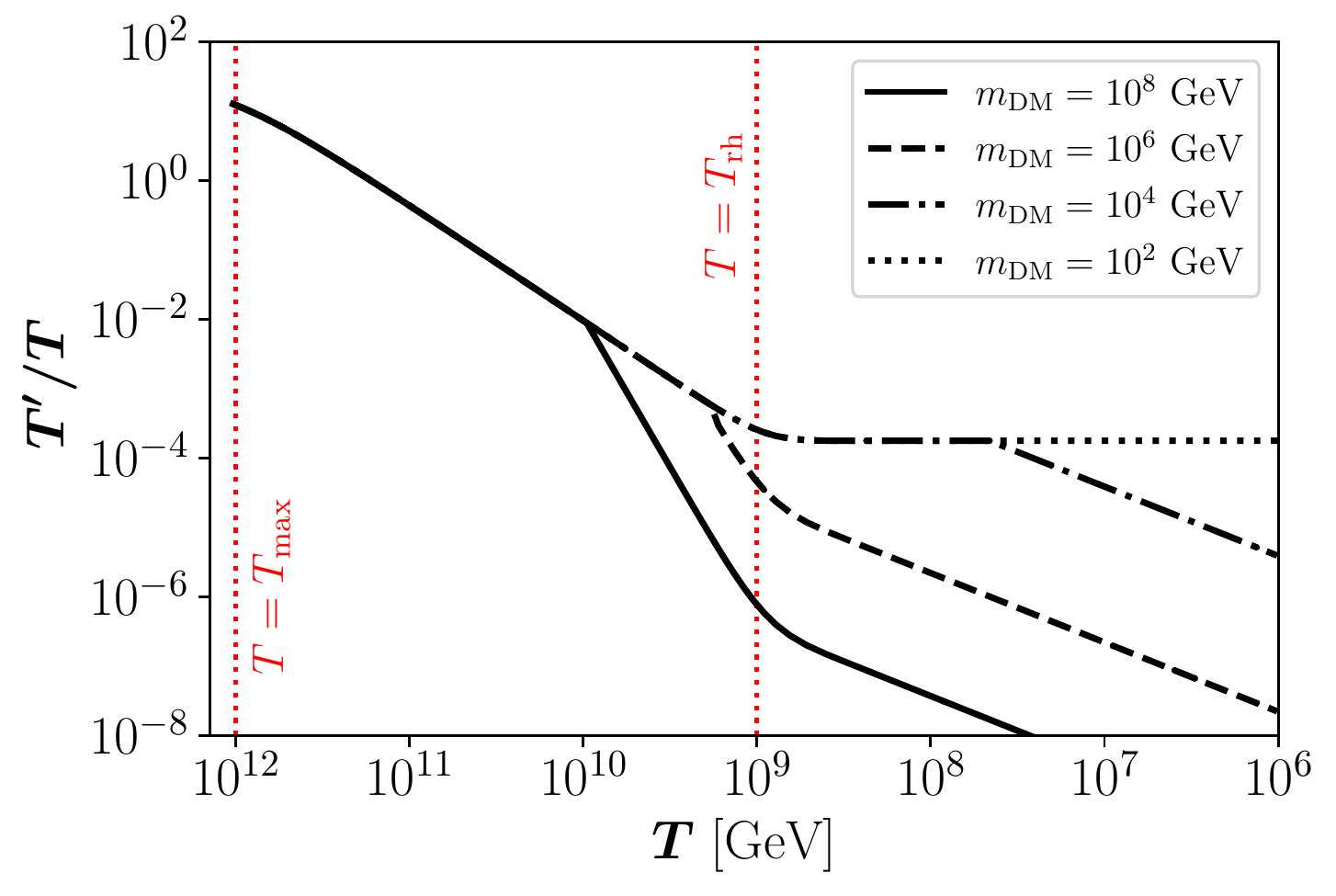}
    \caption{Evolution of the ratio of the DM to SM temperatures $T'/T$ for the case {\it without} chemical equilibrium for different DM masses, taking $\Tmax = 10^{12}$~GeV, $\Trh = 10^9$~GeV and $m_\phi = 3 \times 10^{13}$~GeV.
    }
	\label{fig:DMtemperature}
\end{figure} 
If the dark sector features sizeable self-interactions, DM could reach kinetic equilibrium with itself. As a consequence, the DM distribution is  characterized by a temperature $T'$ which is in general different from the SM temperature $T$. The evolution of the ratio of the DM to SM temperatures $T'/T$ is shown in Fig.~\ref{fig:DMtemperature} for DM produced from the scattering of inflatons. Here we assume $\Trh = 10^9$~GeV, $\Tmax = 10^{12}$~GeV, and inflatons with mass $m_\phi = 3\times 10^{13}$~GeV. During reheating, i.e., while $\Tmax > T > \Trh$, relativistic DM cools down much faster than radiation, $T' \propto a^{-1}$ whereas $T \propto a^{-3/8}$, and therefore $T'/T \propto T^{5/3}$. We note that as inflatons annihilate at rest, initial DM temperature is of the order of inflaton mass. As a result, at $T \simeq \Tmax$ one has  $T'/T \sim m_\phi/\Tmax \sim 30$. After reheating, the relativistic DM and free SM radiation have the same scaling ($T \propto T' \propto a^{-1}$) due to which $T'/T$ remains constant. Eventually, the DM becomes non-relativistic. As it is out-of-chemical equilibrium, its temperature scales as $T' \propto a^{-2}$, and therefore $T'/T \propto T^{13/3}$ or $T'/T \propto T$ for $T \gg \Trh$ or $T \ll \Trh$ respectively. This implies that the DM cools down much faster than the SM. In Fig.~\ref{fig:DMtemperature} we see this nature where the $10^2$~GeV DM is always relativistic for $T \gtrsim 10^6$~GeV. 
Finally, the right panel of Fig.~\ref{fig:DMevolution} shows the equilibrium comoving DM yield (in blue) for different DM masses. For  relativistic DM (i.e., $T' \gg \mdm$), the yield at equilibrium is constant, however it rapidly decreases while becoming nonrelativistic due to  Boltzmann suppression. One may note here, the DM number density is much smaller at production than that at equilibrium.

In addition to kinetic equilibrium, self-interaction can also bring DM into chemical equilibrium with itself. The maximal effect of thermalization and number-changing interactions can be computed following a model-independent approach~\cite{Bernal:2020gzm}. The DM energy density at production, i.e., when $T \simeq \Trh$, can be estimated to be
\begin{equation}
    \rho(\Trh) \simeq Y_0\, s(\Trh)\, \langle E\rangle\,,
\end{equation}
where $\langle E\rangle$ corresponds to the mean DM energy at production.
If DM is produced by annihilation of inflatons, $\langle E\rangle \simeq m_\phi\, (\Trh/\Tmax)^{8/3}$ as the inflatons annihilate at rest.%
\footnote{At $T \simeq \Tmax$, $\langle E\rangle \simeq m_\phi$.
However, $\langle E\rangle \simeq m_\phi\, (\amax/\arh) \simeq m_\phi\, (\Trh/\Tmax)^{8/3}$ at $T = \Trh$.} However, if UV freeze-in is responsible for the DM genesis, $\langle E\rangle \simeq \Trh$ or $\langle E\rangle \simeq \mdm$ for $\mdm \ll \Trh$ or $\mdm \gg \Trh$, respectively.
In the case where DM self-interactions guarantee kinetic equilibrium in the dark sector, DM is characterized by a temperature $T'$ which is in general different from the SM one. Assuming the DM instantaneously thermalizes, its temperature $\Tprh$ at $T = \Trh$ is given by
\begin{equation}
    \Tprh \equiv T'(\Trh) \simeq
    \begin{cases}
        \left[\frac{30}{\pi^2\, \Cr \, g}\, \rho(\Trh)\right]^{1/4} &\text{for }\mdm \ll \Tprh\,,\\[8pt]
        \frac23\, \mdm\, W_0^{-1}\left[\frac{g^{2/3} \mdm^{8/3}}{3\pi\, \rho(\Trh)^{2/3}}\right] &\text{for }\mdm \gg \Tprh\,,
    \end{cases}
\end{equation}
where $\Cr =1$ or $\Cr =7/8$ for bosonic or fermionic DM respectively, $g$ corresponds to the internal DM degrees of freedom and $W_0$ is the $0^\text{th}$ branch of the $W$-Lambert function. Additionally, if DM also reaches chemical equilibrium via number-changing interaction, its equilibrium number density turns out to be
\begin{equation}
    n(\Tprh) =
    \begin{cases}
        \frac{\zeta(3)}{\pi^2}\, \Cn\, g\, \Tprh^3
        &\text{for }\mdm \ll\Tprh\,,\\[8pt]
        \frac{\rho(\Trh)}{\mdm} &\text{for }\mdm \gg \Tprh\,,
    \end{cases}
\end{equation}
where $\Cn =1$ or $3/4$ respectively for bosonic or fermionic DM. Therefore, the maximal yield after thermalization is given by
\begin{equation} \label{eq:Ythermal}
    Y_0^\text{max} = \frac{n(\Tprh)}{s(\Trh)} \simeq
    \begin{cases}
        \frac{15\sqrt{2}\, \Cn\, \zeta(3)}{\pi^4} \left(\frac{3\, g}{\gss}\right)^{1/4} \left[\frac{1}{\Cr}\, \frac{\langle E\rangle}{\Trh}\, Y_0 \right]^{3/4} &\text{for }\mdm \ll \Tprh\,,\\[8pt]
        Y_0\, \frac{\langle E\rangle}{\mdm} &\text{for }\mdm \gg \Tprh\,.
    \end{cases}
\end{equation}

It is important to emphasize that the maximal yield presented in the previous expression may not always be reached. On one hand, DM self-interactions may not be strong enough to guarantee chemical equilibrium, especially for heavy DM, where the unitarity sets upper bounds to the DM self-couplings.
On the other hand, if those couplings are strong enough to make DM number-changing interactions freeze-out when DM is non-relativistic, a cannibalization phase occurs, depleting the DM number density~\cite{Carlson:1992fn, Hochberg:2014dra, Pappadopulo:2016pkp, Farina:2016llk}.

The blue- and red-colored {\it bands} in Fig.~\ref{fig:DM} show the parameter space that generates the entire observed DM abundance once thermalization and number-changing processes are taken into account. The blue bands correspond to the production via scattering of inflatons while UV freeze-in via scattering of the SM particles correspond to the red bands. The bands feature two slopes corresponding to the cases whether the DM is relativistic or not after thermalization. Following Eq.~\eqref{eq:Ythermal}, at high reheating temperatures light DM is produced relativistically, whereas heavy DM at small $\Trh$ is non-relativistic. We observe that, the inclusion of DM self-interaction broadens the allowed parameter space.
It is also intriguing to note that, for fermionic DM with $\Tmax/\Trh \lesssim 20$, DM production via scattering of the SM particles can dominate for masses in the range $10^6$~GeV$\lesssim \mdm \lesssim 10^{10}$~GeV.

In the standard SIMP scenario DM cannibalization leads to a relative rise of the DM temperature with respect to the SM, and therefore poses a potential problem due to the bound on warm DM.
This could be avoided in different ways, for example: $i)$ by demanding kinetic equilibrium between the dark and the visible sectors at the moment of the freeze-out of the number-changing interactions~\cite{Hochberg:2014dra}, $ii)$ by assuming that the dark sector contains additional states that are relativistic when the number-changing interactions freeze-out~\cite{Bernal:2015bla, Bernal:2015lbl}, or $iii)$ by assuming that DM is originally produced much colder than the SM~\cite{Bernal:2015ova, Bernal:2015xba}. Here, however, the bound on warm DM is naturally satisfied since during reheating DM cools down much faster than the SM as the inflaton keeps injecting entropy to the SM thermal bath, cf. Fig.~\ref{fig:DMtemperature}.

\section{Singlet Scalar Dark Matter}\label{sec:ssdm} 
As an example of gravitational DM with sizeable self-interactions, in this section the scalar singlet DM (SSDM) model~\cite{Silveira:1985rk, McDonald:1993ex, Burgess:2000yq} is studied.
In addition to the SM particle spectrum, the SSDM model contains only a real scalar $s$, singlet under the SM gauge group. The scalar $s$ is assumed to be odd under a $\mathbb{Z}_2$ symmetry that guarantees its stability. Thus, the most general renormalizable scalar potential is given by
\begin{equation}
    V = \mu_\mathcal{H}^2\,|\mathcal{H}|^2 + \lambda_\mathcal{H}\,|\mathcal{H}|^4 + \mu_s^2\,s^2 + \lambda_s\,s^4 + \lambda_{\mathcal{H}s}\,|\mathcal{H}|^2\,s^2,
\end{equation}
where $\mathcal{H}$ is the SM Higgs doublet. The phenomenology of this model is completely determined by three parameters
\begin{equation}
    \mdm\,,\, \lambda_{\mathcal{H}s}\,,\, \lambda_s\,,
\end{equation}
corresponding to the DM mass, the Higgs portal and the DM quartic coupling, respectively. One should note here, the role of the DM self-coupling $\lambda_s$ does not influence the DM abundance if it is produced via the standard WIMP or FIMP mechanisms.%
\footnote{It has been very recently pointed out that in this model DM could be also be produced via inverse phase transitions~\cite{Ramazanov:2021eya}.}
However, the self-interaction strength plays a deterministic role in the case where the DM genesis occurs via the SIMP paradigm~\cite{Bernal:2015xba, Heikinheimo:2017ofk, Bernal:2018ins, Almeida:2018oid}.
Due to simplicity and predictability, the phenomenology of SSDM as a WIMP has been widely studied, where it is assumed that the scalar $s$ has a sizeable mixing with the SM Higgs. This scenario has been found to be highly constrained from collider searches~\cite{Barger:2007im, Djouadi:2011aa, Djouadi:2012zc, Damgaard:2013kva, No:2013wsa, Robens:2015gla, Han:2016gyy}, DM direct detection~\cite{He:2009yd, Baek:2014jga, Feng:2014vea, Han:2015hda, Athron:2018hpc} and indirect detection~\cite{Yaguna:2008hd, Goudelis:2009zz, Profumo:2010kp, Cline:2013gha, Urbano:2014hda, Duerr:2015mva, Duerr:2015aka, Benito:2016kyp}. In contrast, scenarios with a very suppressed Higgs portal are much less constrained, and could also lead to a vast phenomenology, such as the DM production via freeze-in~\cite{Yaguna:2011qn, Campbell:2015fra, Kang:2015aqa, Duch:2017khv} or via Hawking evaporation of primordial black holes~\cite{Bernal:2020bjf}. Recent studies have also focused the effect of a non-standard cosmology on this model~\cite{Bernal:2018ins, Hardy:2018bph, Bernal:2018kcw, Allahverdi:2020bys}.

In the singlet scalar model, DM can reach kinetic equilibrium by efficient 2-to-2 elastic scatterings via its quartic interaction (by construction the exchange of a Higgs boson is assumed to be subdominant). The reaction rate for 2-to-2 DM scattering is given by
\begin{equation}
    \Gamma_{2\to 2}(T') = \frac{\lambda_s^2}{64\pi^3} \frac{{T'}^2}{\mdm} \frac{K_1(2\, \mdm/T')}{K_2(\mdm/T')}\,,
\end{equation}
as derived in Appendix~\ref{app:2to2}. Apart from kinetic equilibrium, chemical equilibrium is also required to have reproductive DM interactions. In the singlet scalar DM model, this comes mainly from 4-to-2 scatterings, the 3-to-2 processes being forbidden by the $\mathbb{Z}_2$ symmetry. The interaction rate depends on the quartic coupling $\lambda_s$, and is estimated to be
\begin{equation}
    \Gamma_{2 \to 4}(T') \simeq
    \begin{cases}
        2.4\times 10^{-5}\, \lambda_s^4\, T' & \text{ for } \mdm \ll T',\\[8pt]
        \frac{27}{128\, \pi^{11/2}} \sqrt{\frac32}\, \lambda_s^4\, T' \left(\frac{T'}{\mdm}\right)^{7/2} e^{-3\frac{\mdm}{T'}} & \text{ for } \mdm \gg T',
    \end{cases}
\end{equation}
as reported in Appendix~\ref{app:2to4}. It is worth mentioning here that the self-interaction strength $\lambda_s$ is bounded from above from the requirement of perturbative unitarity~\cite{Gonderinger:2009jp, Lerner:2009xg, Ko:2014nha} that necessarily demands $\lambda_s < 4\pi$.

In order to account for the evolution of DM number density with self-interaction we include both the gravitational production and number-changing processes corresponding to 2-to-4 and 4-to-2 reactions to the BEQ in Eq.~\eqref{eq:BE0}
\begin{equation}
    \frac{dn}{dt} + 3H\,n = \gamma + \gamma_{2\to 4} \left(\frac{n}{n_\text{eq}}\right)^2\, \left[1 - \left(\frac{n}{n_\text{eq}}\right)^2\right],
\end{equation}
where $\gamma_{2\to 4} \equiv n_\text{eq}\, \Gamma_{2\to 4} \equiv n_\text{eq}^2\, \langle\sigma v\rangle_{2\to 4}$ is the 2-to-4 rate density, and $n_\text{eq}(T')$ is the equilibrium DM number density.

\begin{figure}
    \def\sepf{0.50}
	\centering
	\includegraphics[scale=\sepf]{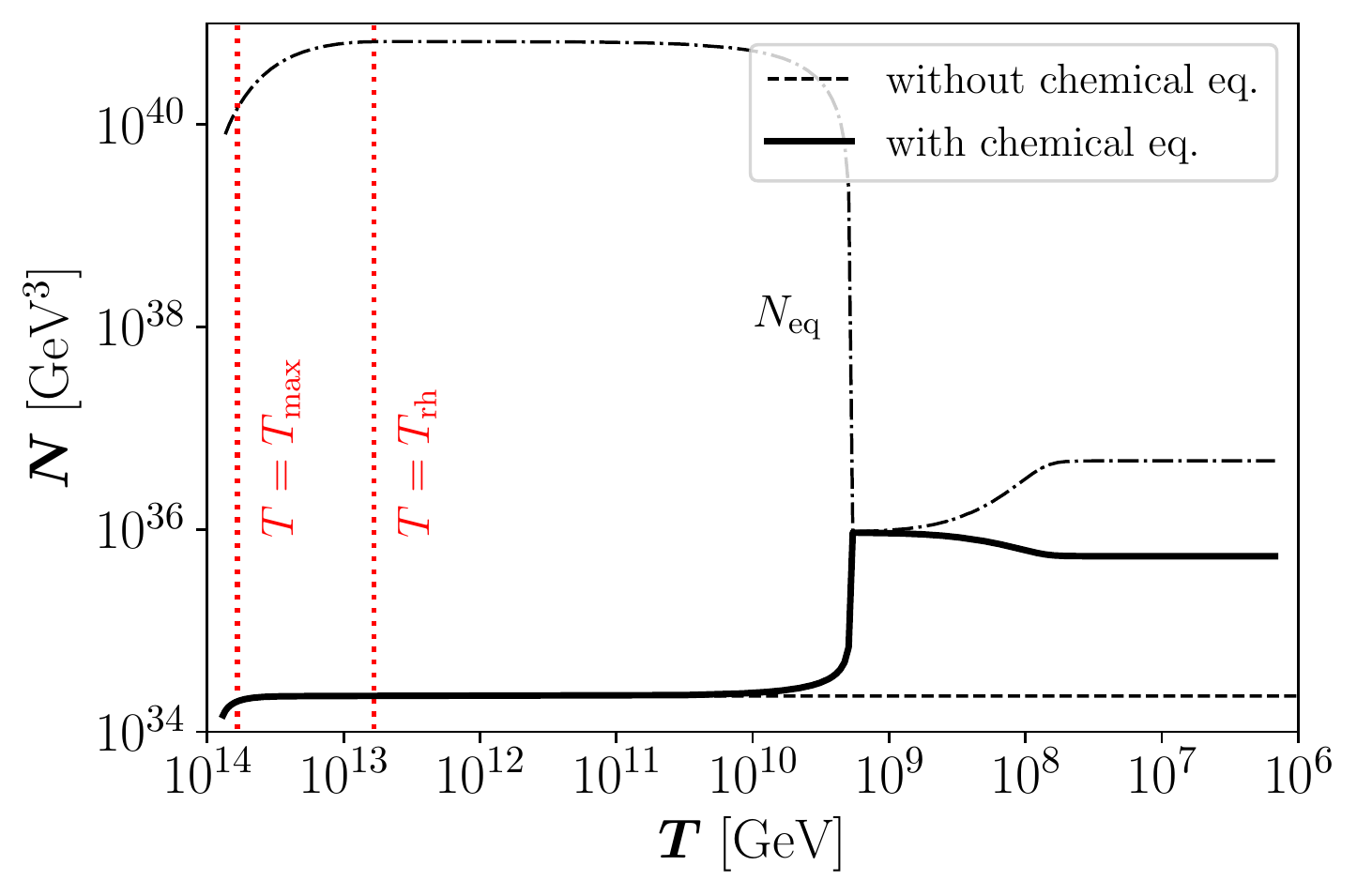}
	\includegraphics[scale=\sepf]{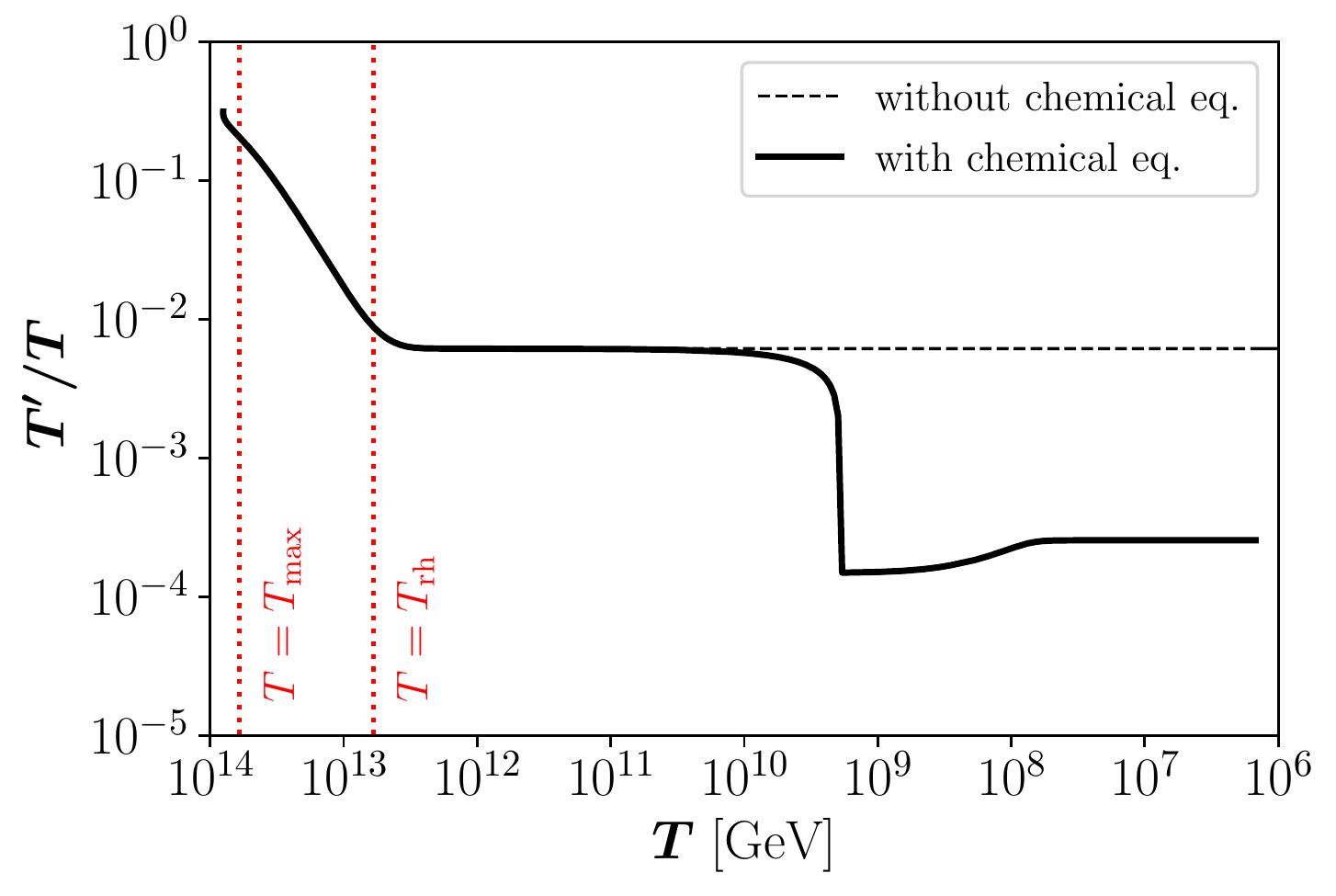}
    \caption{Evolution of the comoving DM yield $N$ (left panel) and the ratio of temperatures $T'/T$ (right panel) as a function of the SM temperature $T$ for the SSDM model for $\Trh = 6\times 10^{12}$~GeV and $\Tmax/\Trh = 10$, assuming $m_\phi = 3 \times 10^{13}$~GeV.
    We have also taken $\mdm = 8 \times 10^4$~GeV, and $\lambda_s = 10$.
    The vertical red dotted lines correspond to $T = \Tmax$ and $T = \Trh$.
    The solid thick black lines correspond to the case with chemical equilibrium and yields the observed DM abundance, whereas the dashed black lines correspond to the case without chemical equilibrium (with $\lambda_s\to 0$) that gives rise to an underabundant DM density. The black dot-dashed line in the left panel plot denotes $N_\text{eq}(T')$ corresponding to the equilibrium DM number density. 
    }
	\label{fig:DMevolutionWITH}
\end{figure} 

Figure~\ref{fig:DMevolutionWITH} shows the evolution of the comoving DM yield $N$ (left panel) and the ratio of temperatures $T'/T$ (right panel) as a function of the SM temperature $T$ for $\Trh = 6 \times 10^{12}$~GeV and $\Tmax/\Trh = 10$, assuming $m_\phi = 3\times 10^{13}$~GeV.
We have also considered $\mdm = 8 \times 10^4$~GeV and $\lambda_s = 10$. The vertical red dotted lines correspond to $T = \Tmax$ and $T = \Trh$. The solid thick black curves denote the scenario with chemical equilibrium that gives rise to the observed DM abundance, while the black dashed lines indicate the case without chemical equilibrium (similar to Fig.~\ref{fig:DMevolution} and~\ref{fig:DMtemperature}) that generate a DM underabundance. The evolution in the case where chemical equilibrium is reached contrasts with the one without in two main aspects:
firstly, there is a fast increase of the DM number density due to the out-of-equilibrium 2-to-4 processes at the cost of reducing its temperature. This period ends at $T \simeq 2\times 10^9$~GeV when chemical equilibrium is reached as seen from the left panel of Fig.~\ref{fig:DMevolutionWITH} where the equilibrium number density of the DM is shown by the black dot-dashed curve. Secondly, once the DM becomes non-relativistic, 4-to-2 cannibalizations become effective. This causes the DM number density to reduce and the DM temperature falls less rapidly since $T' \propto 1/\log a$~\cite{Carlson:1992fn}. As a result, the DM is heated up with respect to the SM and the ratio $T'/T$ rises at $T \simeq 3\times 10^8$~GeV. The number-changing interactions eventually freeze out at $T \simeq 7\times 10^7$~GeV, breaking the chemical equilibrium.

Finally, we note that DM self-interactions are bounded. In fact, Bullet cluster measurements point towards DM self-interactions below 1.25~cm$^2$/g at $68\%$ CL~\cite{Markevitch:2003at, Clowe:2003tk, Randall:2007ph}.
Moreover, recent observational data on cluster collisions have led to a more stringent bound on $\sigma_{2\to 2}/\mdm < 0.47$~cm$^2$/g at 95\% CL~\cite{Harvey:2015hha}, where for the SSDM model
\begin{equation}
    \frac{\sigma_{2\to 2}}{\mdm} \simeq \frac{9}{8\, \pi} \frac{\lambda_s^2}{\mdm^3}\,.
\end{equation}
However, this bound constrains sub-GeV DM and not reachable in this context of gravitational production.

\section{Conclusions}\label{sec:con}
Dark matter (DM) produced from inflaton scattering mediated by massless gravitons during the era of inflation can account for the entire observed DM abundance ranging from DM masses of a few TeV to as large as a few PeV depending on the choice of the reheating temperature, and the maximum temperature achieved during reheating. At the same time, DM can also be produced from 2-to-2 annihilation of the Standard Model (SM) bath particles via graviton exchange leading to the standard UV freeze-in mechanism. DM production via such gravity mediated processes is unavoidable and can yield the whole observed relic density. Now, even if gravity remains as the {\it only} portal between the dark and the visible sector, the DM particles still can have sizeable interactions among themselves which can lead to kinetic equilibrium within the  dark sector. In that case, the dark sector can evolve with a distribution that is characterized by a temperature generally different from the SM bath. On top of kinetic equilibrium, number-changing interactions can result in chemical equilibrium of the DM with itself.

In the present work, by considering the thermalization due to number-changing interactions within the dark sector, we have shown that the gravitationally produced DM can give rise to the observed relic abundance over a broader region of the parameter space which is otherwise absent. The allowed parameter space remains safe from the Lyman-$\alpha$ bound on warm DM mass but conflicts with CMB measurements on the upper limit on the inflationary scale for large reheating temperatures. The analysis is model independent and applies to DM particles with 0, 1/2, or 1 intrinsic spin. Although the gravitational DM production from the SM bath is generally subdominant, in the presence of self-interactions it could dominate for fermionic DM in the mass window of $10^6$ to $10^{10}$~GeV for $\Tmax/\Trh\lesssim 20$.    

We finally take up the $\mathbb{Z}_2$ symmetric scalar singlet DM model as an example of a minimal extension of the SM where self-interaction amongst the DM particles can arise naturally. In this case, the DM reaches kinetic equilibrium via 2-to-2 scattering process which is controlled by the quartic interaction strength. The chemical equilibrium, on the other hand, is achieved via the number-changing 2-to-4 and 4-to-2  processes. Considering the gravitational production of the scalar DM, to obtain the yield, we numerically solve a set of coupled Boltzmann equations by introducing DM self-interactions. We find, for large enough quartic coupling satisfying perturbative unitarity, the scalar DM can produce the required abundance by remaining well within chemical equilibrium.

\section*{Acknowledgments}
The authors would like to acknowledge discussions with Xiaoyong Chu and Purusottam Ghosh. 
NB received funding from Universidad Antonio Nariño grants 2019101, and 2019248, the Spanish FEDER/MCIU-AEI under grant FPA2017-84543-P, and the Patrimonio Au\-tó\-no\-mo - Fondo Nacional de Financiamiento para la Ciencia, la Tecnología y la Innovación Francisco José de Caldas (MinCiencias - Colombia) grant 80740-465-2020.
This project has received funding/support from the European Union's Horizon 2020 research and innovation programme under the Marie Skłodowska-Curie grant agreement No 860881-HIDDeN.

\appendix

\section{Bound on the Reheating Temperature} \label{sec:Trh}
The evolution of the inflaton and SM radiation energy densities can be tracked via the set of Boltzmann equations given in Eqs.~\eqref{eq:BErhop} and~\eqref{eq:BErhoR}.
During reheating, the inflaton dominates the total energy density (i.e., $H^2 \simeq \rho_\phi / (3M_P^2)$) and evolves as
\begin{equation} \label{eq:rhophi}
    \rho_\phi(a) = \rho_\phi(a_I) \left[\frac{a_I}{a}\right]^3 = 3M_P^2\, H_I^2 \left[\frac{a_I}{a}\right]^3,
\end{equation}
with $a$ the scale factor and $H_I \equiv H(a_I)$, where $a_I$ corresponds to the scale factor at the beginning of the reheating era.
The evolution of the radiation energy density can be extracted from Eqs.~\eqref{eq:BErhoR} and~\eqref{eq:rhophi}, and reads
\begin{equation}
    \rho_R(a) = \frac65\, M_P^2\, \Gamma_\phi\, H_I\, \left(\frac{a_I}{a}\right)^4\, \left[\left(\frac{a}{a_I}\right)^{5/2}-1\right].
\end{equation}
It follows that the thermal bath reaches a maximum temperature $T = \Tmax$ when only a small fraction of the inflaton has decayed~\cite{Chung:1998rq, Giudice:2000ex} at $a = a_\text{max} \equiv (8/3)^{2/5}\, a_I$ that corresponds to
\begin{equation} \label{eq:Tmax}
    \Tmax^4 = \frac{60}{\pi^2\,\gs} \left(\frac38\right)^{8/5} M_P^2\, \Gamma_\phi\, H_I\, .
\end{equation}
Finally, the Hubble expansion rate at the beginning of the heating era takes the form~\cite{Bernal:2019mhf}
\begin{equation}
    H_I = \frac{\pi}{2} \sqrt{\frac{\gs}{10}} \left(\frac83\right)^{8/5} \frac{\Tmax^4}{M_P\, \Trh^2}\,,
\end{equation}
which, taking into account the upper limit on the inflationary scale $H_I^\text{CMB} \leq 2.5\times 10^{-5}~M_P$ \cite{Akrami:2018odb}, allows to extract an upped bound on $\Trh$:
\begin{equation} \label{eq:boundHI}
    \Trh \leq \sqrt{\frac{2}{\pi}} \left(\frac{10}{\gs}\right)^\frac14 \left(\frac38\right)^\frac45 \left(\frac{\Trh}{\Tmax}\right)^2 \sqrt{M_P\, H_I^\text{CMB}}
    \lesssim 2.5\times 10^9~\text{GeV}\left[\frac{10^3}{\Tmax/\Trh}\right]^2.
\end{equation}

\section{\boldmath Lyman-$\alpha$ Constraint}\label{app:lyman}
Due to their large initial momentum, DM particles could have a large free-streaming length leading to a suppression on the structure formation at small scales.
In the present scenario where DM has no interactions with the SM or with itself, the DM momentum simply redshifts, and its value $p_0$ at present is~\cite{Fujita:2014hha}
\begin{equation}
    p_0 = \frac{a_\text{in}}{a_0}\, p_\text{in}
    = \frac{a_\text{in}}{a_\text{eq}} \frac{\Omega_R}{\Omega_m}\, p_\text{in} =
    \begin{cases}
        \left[\frac{\gss(T_\text{eq})}{\gss(T_\text{in})}\right]^{1/3} \frac{T_\text{eq}}{T_\text{in}} \frac{\Omega_R}{\Omega_m}\, p_\text{in} \quad &\text{for } T_\text{in} \leq \Trh \,,\\[8pt]
        \left[\frac{\gss(T_\text{eq})}{\gss(\Trh)}\right]^{1/3} \left(\frac{\Trh}{T_\text{in}}\right)^{8/3} \frac{T_\text{eq}}{\Trh} \frac{\Omega_R}{\Omega_m}\, p_\text{in} \quad &\text{for } T_\text{in} \geq \Trh \,,
    \end{cases}
\end{equation}
where $p_\text{in}$ is the mean initial momentum at production ($a = a_\text{in}$), $T_\text{eq}$ and $a_\text{eq}$ correspond to the temperature and the scale factor at the matter-radiation equality, respectively.
For DM produced by the scattering of SM particles $T_\text{in} \simeq p_\text{in} \simeq \Trh$, and
\begin{equation}
    p_0 \simeq
    \left[\frac{\gss(T_\text{eq})}{\gss(\Trh)}\right]^{1/3} \frac{\Omega_R}{\Omega_m}\, T_\text{eq}\,.
\end{equation}
In this case, the usual lower bound from warm DM applies, corresponding to $\mdm \simeq 3.5$~keV~\cite{Irsic:2017ixq}.
However, if DM is produced by the scattering of inflatons then $T_\text{in} \simeq \Tmax$ and $p_\text{in} \simeq m_\phi$, therefore the typical momentum for DM at present epoch can be estimated as
\begin{equation}
    p_0 \simeq \left[\frac{\gss(T_\text{eq})}{\gss(\Trh)}\right]^{1/3} \left(\frac{\Trh}{\Tmax}\right)^{8/3} \frac{T_\text{eq}}{\Trh} \frac{\Omega_R}{\Omega_m}\, m_\phi\,.
\end{equation}
A lower bound on the DM mass can be obtained from the upper bound on a typical velocity of warm DM at the present time.
Taking $v_\text{DM} \lesssim 1.8 \times 10^{-8}$~\cite{Masina:2020xhk} for $\mdm \simeq 3.5$~keV~\cite{Irsic:2017ixq}, one gets
\begin{equation}
    \mdm \gtrsim 0.8~\text{eV} \left(\frac{10^3}{\Tmax/\Trh}\right)^{8/3} \frac{10^9~\text{GeV}}{\Trh}\, \frac{3\times 10^{13}~\text{GeV}}{m_\phi}\,,
\end{equation}
where $T_\text{eq} \simeq 0.8$~eV, $\Omega_R \simeq 5.4 \times 10^{-5}$ and $\Omega_m \simeq 0.315$~\cite{Aghanim:2018eyx, Zyla:2020zbs} were used.
It is interesting to note that in this case DM in the eV ballpark can be explored, easing the tension with structure formation.
In fact, DM naturally tends to be much colder than the SM due to the fact that during reheating SM radiation is not free and scales as $T \propto a^{-3/8}$ whereas DM momentum simply redshifts.

\section{Relevant Vertices}\label{sec:vertex}
In Tab.~\ref{tab:vert-fac} we have collected the relevant vertex factors~\cite{Choi:1994ax} for graviton-DM interaction for the case of spin-0 $S$, spin-1/2 $\psi$, and spin-1 $X_\mu$ DM. 
\FloatBarrier
\begin{table}[htb!]\scriptsize
\begin{center}
\begin{tabular}{|c||c|}
\hline
Interaction Vertex  &  Vertex factors\\
\hline\hline
&\\
\begin{minipage}{.15\textwidth}
      \includegraphics[scale=0.3]{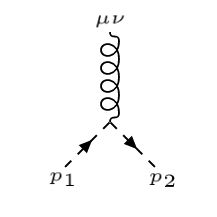}
    \end{minipage}& $-\frac{i}{2M_P}\left[p_{1\mu}p_{2\nu}+p_{1\nu}p_{2\mu}-\eta_{\mu\nu}\left(p_1 \cdot p_2-m_S^2\right)\right]$  \\\cline{2-2}
    &\\
\begin{minipage}{.15\textwidth}
      \includegraphics[scale=0.3]{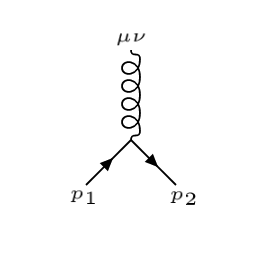}
    \end{minipage}& $-\frac{i}{8M_P}\left[\left(p_1+p_2\right)_\mu\gamma_\nu+\left(p_1+p_2\right)_\nu\gamma_\mu-2\eta_{\mu\nu}\left(\slashed{p_1}+\slashed{p_2}-2m_\psi\right)\right]$  \\\cline{2-2}
    &\\
\begin{minipage}{.15\textwidth}
      \includegraphics[scale=0.3]{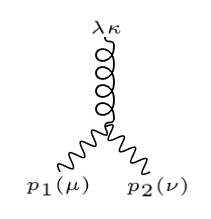}
    \end{minipage} & $\begin{aligned}
-\frac{i}{2M_P}\Bigl[\eta_{\lambda\kappa}\eta_{\mu\nu}\left(p_1\cdot p_2-m_X^2\right)
    -\eta_{\lambda\kappa}p_{1\nu}p_{2\mu}+\eta_{\kappa\mu}p_{1\nu}p_{2\lambda}-\eta_{\mu\nu}p_{1\kappa}p_{2\lambda}&\\ +\eta_{\lambda\nu}p_{1\kappa}p_{2\mu}-\eta_{\kappa\mu}\eta_{\lambda\nu}\left(p_1\cdot p_2-m_X^2\right)-\eta_{\mu\nu}p_{1\lambda}p_{2\kappa}+\eta_{\lambda\mu}p_{1\nu}p_{2\kappa}\\-\eta_{\kappa\nu}\eta_{\lambda\mu}\left(p_1 \cdot p_2-m_X^2\right)\Bigr]\end{aligned}$ \\ 
 &\\
 \hline
\end{tabular}
\end{center}
\caption {Relevant graviton-DM vertices for scalar $(S)$, fermion $(\psi)$ and vector boson $(X)$ DM, from top to bottom. For the inflaton-graviton interaction one has to replace $m_S$ with $m_\phi$.}
\label{tab:vert-fac}
\end{table}\FloatBarrier

\section{Dark Matter Production Rates from Inflaton Annihilations} \label{sec:rates}
For 2-to-2 process we can define the following kinematics in the center-of-mass (CM) frame for a $1,2\to3,4$ scattering considering all states to be massive:
\begin{equation}
    \begin{split}
        &E_{\text{1CM}}=\frac{s+m_1^2-m_2^2}{2\sqrt{s}},~E_{\text{2CM}}=\frac{s+m_2^2-m_1^2}{2\sqrt{s}},\\
        & E_{\text{3CM}}=\frac{s+m_3^2-m_4^2}{2\sqrt{s}}, E_{\text{4CM}}=\frac{s+m_4^2-m_3^2}{2\sqrt{s}},
    \end{split}\label{eq:kinemat}
\end{equation}
where $\sqrt{s}$ is the CM energy for the process. Now, the relevant amplitude for $\phi,\phi\to\text{DM},\text{DM}$ is given by 
\begin{equation}
    \mathcal{M}^{\phi j}\propto \mathcal{M}_\phi^{\mu\nu}\Pi^{\mu\nu\rho\sigma}\mathcal{M}_{\rho\sigma}^j 
\end{equation}
where $j$ denotes the spin of the DM involved in the production process.
The graviton propagator is defined via
\begin{equation}
    \Pi^{\mu\nu\rho\sigma}=\frac{1}{2k^2}\Bigl(\eta^{\rho\nu}\eta^{\sigma\mu}+\eta^{\rho\mu}\eta^{\sigma\nu}-\eta^{\rho\sigma}\eta^{\mu\nu}\Bigr), 
\end{equation}
with $k=p_1+p_2=\sqrt{s}$ as the sum of the four-momenta of the initial state (inflatons). We can write the partial amplitudes following Tab.~\ref{tab:vert-fac} as
\begin{equation}
    \begin{split}
        &\mathcal{M}_{\phi(S)}^{\mu\nu}=\frac{1}{2}\left(p_{1(3)\mu}p_{2(4)\nu}+p_{1(3)\nu}p_{2(4)\mu}-\eta_{\mu\nu}p_{1(3)}\cdot p_{2(4)}-\eta_{\mu\nu}m_{\phi(S)}^2\right)\\
        & \mathcal{M}_{\psi}^{\mu\nu}= \frac{1}{4}\,\overline{v}\left(p_4\right)\Bigl[\gamma_\mu\left(p_3-p_4\right)_\nu+\gamma_\nu\left(p_3-p_4\right)_\mu\Bigr]u\left(p_3\right)\\&
        \mathcal{M}_X^{\mu\nu}=\frac{1}{2}\Biggl(\left(p_3\cdot p_4-m_X^2\right) \eta^{\mu \nu } \eta^{\rho \sigma }-\left(p_3\cdot p_4-m_X^2\right) \eta^{\mu \sigma } \eta^{\nu \rho }-\left(p_3\cdot p_4-m_X^2\right) \eta^{\mu \rho } \eta^{\nu \sigma }\\&-p_3^{\sigma } p_4^{\rho } \eta^{\mu \nu }+p_3^{\sigma } p_4^{\nu } \eta^{\mu \rho }+p_3^{\nu } p_4^{\rho } \eta^{\mu \sigma }+p_3^{\sigma } p_4^{\mu } \eta^{\nu \rho }+p_3^{\mu } p_4^{\rho } \eta^{\nu \sigma }-p_3^{\nu } p_4^{\mu } \eta^{\rho \sigma }-p_3^{\mu } p_4^{\nu } \eta^{\rho \sigma }\Biggr)\epsilon\left(p_3\right)_{\rho}^{\star}\epsilon\left(p_4\right)_{\sigma}^{\star},
    \end{split}
\end{equation}
where $p_{3,4}$ is the four momenta of the outgoing states (DM). The 2-to-2 cross-section in the CM frame can be expressed as
\begin{equation}
    \begin{split}
        \Bigl(\sigma v_\text{rel}\Bigr)_{\phi\phi\to\text{DM}\text{DM}}=\frac{1}{64\pi^2 m_\phi^2}\frac{\left|\vec{p_f}\right|}{\sqrt{s}}\int \overline{\mathcal{M}^2}_{\phi\phi\to\text{DM}\text{DM}}d\Omega
    \end{split}\label{eq:sigv}
\end{equation}
with
\begin{equation}
    \left|\vec{p_f}\right| = \frac{\sqrt{s}}{2}\sqrt{1-\frac{4\mdm^2}{s}}\label{eq:pf}
\end{equation}
as the three-momentum of the final state. Substituting Eq.~\eqref{eq:pf} in Eq.~\eqref{eq:sigv} we then find
\begin{equation}\label{eq:sigv2}
    \Bigl(\sigma v_\text{rel}\Bigr)_{\phi\phi\to\text{DM}\text{DM}}=\frac{\overline{\mathcal{M}^2}_{\phi\phi\to\text{DM}\text{DM}}}{32\pi m_\phi^2}\sqrt{1-\frac{\mdm^2}{m_\phi^2}}\, .
\end{equation}

\subsection{Scalar Dark Matter}
For the scalar DM case $S$, using Eq.~\eqref{eq:kinemat} we can write the spin-averaged squared amplitude as
\begin{equation}
    \overline{\mathcal{M}^2}_{\phi\phi\to SS}=\frac{1}{64M_P^4}\left(m_S^2+2m_\phi^2\right)^2   
\end{equation}
which on substitution in Eq.~\eqref{eq:sigv2} gives rise to
\begin{equation}
    \Bigl(\sigma v_\text{rel}\Bigr)_{\phi\phi\to SS}=\frac{m_\phi^2}{4096\pi M_P^4}\left(x^2+2\right)^2\sqrt{1-x^2}
\end{equation}
with $x=m_S/m_\phi$. Corresponding DM production rate is then given by
\begin{equation}
    \Gamma_{\phi\phi\to SS}=n_\phi\Bigl(\sigma v_\text{rel}\Bigr)_{\phi\phi\to SS}=\frac{\rho_\phi m_\phi}{4096\pi M_P^4}\left(x^2+2\right)^2\sqrt{1-x^2}\,.
\end{equation}

\subsection{Fermionic Dark Matter}
For the Dirac fermionic DM case $\psi$, using Eq.~\eqref{eq:kinemat} , we obtain
\begin{equation}
    \overline{\mathcal{M}^2}_{\phi\phi\to \psi\psi} = \frac{m_\chi^2}{128 M_P^4}\left(m_\phi^2-m_\chi^2\right)
\end{equation}
which in turn, following Eq.~\eqref{eq:sigv2}, gives rise to
\begin{equation}
    \Bigl(\sigma v_\text{rel}\Bigr)_{\phi\phi\to\psi\psi}= \frac{m_\phi^2}{4096\pi M_P^4}x^2\left(1-x^2\right)^{3/2}
\end{equation}
where $x=m_\psi/m_\phi$. Then the rate of fermionic DM turns out to be 
\begin{equation}
    \Gamma_{\phi\phi\to\psi\psi}=n_\phi \Bigl(\sigma v_\text{rel}\Bigr)_{\phi\phi\to\psi\psi}=\frac{\rho_\phi m_\phi}{4096\pi M_P^4}x^2\left(1-x^2\right)^{3/2}. 
\end{equation}

\subsection{Vector Dark Matter}
The spin-averaged amplitude squared for the case of massive vector DM $X$ reads
\begin{equation}
    \overline{\mathcal{M}^2}_{\phi\phi\to XX}=\frac{m_\phi^4}{1024 M_P^4}\,\Bigl(4+ 4 x^2 + 19 x^4\Bigr)
\end{equation}
Following the CM frame kinematics as before we get the 2-to-2 annihilation cross-section as 
\begin{equation}
    \Bigl(\sigma v_\text{rel}\Bigr)_{\phi\phi\to XX}=\frac{  m_\phi^2}{32768\pi M_P^4}\sqrt{1-x^2}\,\Bigl(4+ 4 x^2 + 19 x^4\Bigr), 
\end{equation}
where $x=m_X/m_\phi$. The rate of DM production is given as
\begin{equation}
    \Gamma_{\phi\phi\to XX}=\frac{\rho_\phi m_\phi}{32768\pi M_P^4}\sqrt{1-x^2}\,\Bigl(4+ 4 x^2 + 19 x^4\Bigr).
\end{equation}

\section{Rates for Dark Matter Self-interaction}
In this section, we compute the 2-to-2 and 2-to-4 DM interaction rates for the case of the SSDM model, in the limit where $\lambda_{\mathcal{H}s} = 0$.

\subsection{2-to-2 Processes} \label{app:2to2}
The thermally averaged 2-to-2 elastic-scattering cross-section is given by~\cite{Gondolo:1990dk}
\begin{equation}\label{eq:sigv22mass}
    \langle\sigma v\rangle_{2\to2}(T') = \frac{1}{8\mdm^4\, T'\, K_2\left(\mdm/T'\right)^2}\int_{4\mdm^2}^\infty ds\, \sigma\left(s\right)\left(s-4\mdm^2\right)\sqrt{s}\, K_1\left(\frac{\sqrt{s}}{T'}\right),
\end{equation}
where $\sigma\left(s\right)=\lambda_s^2/32\pi s$. This leads to the 2-to-2 interaction rate
\begin{equation}
    \Gamma_{2\to 2}(T')=n_\text{eq}\, \langle\sigma v\rangle_{2\to2} = \frac{\lambda_s^2}{64\pi^3} \frac{{T'}^2}{\mdm} \frac{K_1(2\, \mdm/T')}{K_2(\mdm/T')}\,. 
\end{equation}

\subsection{2-to-4 Processes} \label{app:2to4}
In the nonrelativistic limit, the thermally averaged 4-to-2 cross-section can be expressed as~\cite{Bernal:2015xba}
\begin{equation}\label{eq:sigv42}
    \langle\sigma v^3\rangle_{4\to2} \simeq \frac{\sqrt{3}}{256\pi \mdm^4}\left|\mathcal{M}\right|_{4\to2}^2 \simeq \frac{27\sqrt{3}}{8\pi}\frac{\lambda_s^4}{\mdm^8}\, .
\end{equation}
Thus, the nonrelativistic 2-to-4 equilibrium rate turns out to be
\begin{equation} \label{eq:rate24nonrel}
    \Gamma_{2\to4}(T') = n_\text{eq}^3 \, \langle\sigma v^3\rangle_{4\to2} \simeq \frac{27}{128\,\pi^{11/2}}\sqrt{\frac32}\, \lambda_s^4\, T' \left(\frac{T'}{\mdm}\right)^{7/2} e^{-3\frac{\mdm}{T'}}.   
\end{equation}

Alternatively, in the ultrarelativistic limit, the 2-to-4 cross-section behaves like
\begin{equation}\label{eq:sigv24rel}
    \sigma_{2\to4}\left(s\right) \simeq \mathcal{K}\, \frac{\lambda_s^4}{s}\,,
\end{equation}
where $\mathcal{K} \simeq 2\times 10^{-4}$ is a dimensionless prefactor numerically computed with {\tt Calchep}~\cite{Belyaev:2012qa}.
The interaction rate in the ultrarelativistic limit can be written as
\begin{equation}\label{eq:rate24rel}
    \Gamma_{2\to4}(T') = n_\text{eq}\, \langle\sigma v\rangle_{2\to4}
    \simeq \frac{\zeta(3)}{\pi^2}\, \mathcal{K}\, \lambda_s^4\, T'\,.
\end{equation}
Equations~\eqref{eq:rate24nonrel} and~\eqref{eq:rate24rel} match when $T' \simeq \mdm$.

\bibliographystyle{JHEP}
\bibliography{biblio}
\end{document}